\documentclass[english,aps,pre,reprint,superscriptaddress,hidelinks]{revtex4-2}
\usepackage[T1]{fontenc}
\usepackage[utf8]{inputenc}
\usepackage{mathtools}
\usepackage{amsmath}
\usepackage{amsthm}
\usepackage{amssymb}
\usepackage{graphicx}

\makeatletter
\theoremstyle{plain}
\newtheorem{thm}{\protect\theoremname}
\theoremstyle{definition}

\usepackage[utf8]{inputenc}
\usepackage{amsmath}

\DeclareMathOperator*{\argmax}{\arg\!\max}
\usepackage{newtxtext}

\makeatother

\usepackage{babel}
\providecommand{\examplename}{Example}
\providecommand{\theoremname}{Theorem}

\begin{document}
\title{Partial information decomposition: redundancy as information bottleneck}
\author{Artemy Kolchinsky}
\email{artemyk@gmail.com}

\affiliation{ICREA-Complex Systems Lab, Universitat Pompeu Fabra, 08003 Barcelona,
Spain}
\affiliation{Universal Biology Institute, The University of Tokyo, 7-3-1 Hongo,
Bunkyo-ku, Tokyo 113-0033, Japan}
\begin{abstract}
The partial information decomposition (PID) aims to quantify the amount
of redundant information that a set of sources provides about a target.
Here, we show that this goal can be formulated as a type of information
bottleneck (IB) problem, termed the ``redundancy bottleneck''
(RB). The RB formalizes a tradeoff between prediction and compression:
it extracts information from the sources that best predict the target,
without revealing which source provided the information. It can be
understood as a generalization of ``Blackwell redundancy'', which we
previously proposed as a principled measure of PID redundancy. The
``RB curve'' quantifies the prediction--compression tradeoff at multiple
scales. This curve can also be quantified for individual sources,
allowing subsets of redundant sources to be identified without combinatorial
optimization. We provide an efficient iterative algorithm for computing
the RB curve.
\end{abstract}
\maketitle
\global\long\def\vvss{s}%
\global\long\def\sdist{\nu_{S}}%
\global\long\def\rrate{R}%
\global\long\def\RR{I_{\cap}}%
\global\long\def\RBbase{I_{\text{RB}}}%
\global\long\def\RB#1{\RBbase(#1)}%
\global\long\def\RBR{\RB R}%

\global\long\def\icons{\chi}%

\global\long\def\DDbase#1{V(#1)}%
\global\long\def\DD{\DDbase{\icons}}%
\global\long\def\where{\quad\textrm{where}\quad}%
\global\long\def\ibL{\beta}%
\global\long\def\cebL{\beta}%
\global\long\def\eqname{Eq.~}
\global\long\def\eqsname{Eqs.~}

\section{Introduction}

Many research fields that study complex systems are faced with multivariate
probabilistic models and high-dimensional datasets. Prototypical examples
include brain imaging data in neuroscience, gene expression data in
biology, and neural networks in machine learning. In response, various
information-theoretic frameworks have been developed in order to study
multivariate systems in a universal manner. Here, we focus on two such
frameworks, \emph{partial information decomposition} and the\emph{
information bottleneck.}

The \emph{partial information decomposition} (PID) considers how information
about a target random variable $Y$ is distributed among a set of
source random variables $X_{1},\dots,X_{n}$ \citep{williams2010nonnegative,wibral_partial_2017,lizier2018information,kolchinskyNovelApproachPartial2022}.
For example, in neuroscience, the sources $X_{1},\dots,X_{n}$ might
represent the activity of $n$ different brain regions and $Y$ might
represent a stimulus, and one may wish to understand how information
about the stimulus is encoded in different brain regions. A central
idea of the PID is that the information provided by the sources can
exhibit \emph{redundancy}, when the same information about $Y$ is
present in each source, and \emph{synergy}, when information about
$Y$ is found only in the collective outcome of all sources.  Moreover,
it has been shown that standard information-theoretic quantities,
such as entropy and mutual information, are not sufficient to quantify
redundancy and synergy \citep{williams2010nonnegative,williams_information_2011}.
However, finding the right measures of redundancy and synergy has
proven difficult. In recent work \citep{kolchinskyNovelApproachPartial2022},
we showed that such measures can be naturally defined by formalizing
the analogy between set theory and information theory that lies at
the heart of the PID \citep{williams_information_2011}. We then proposed
a measure of redundant information (\emph{Blackwell redundancy})
that is motivated by algebraic, axiomatic, and operational considerations.
We argued that Blackwell redundancy overcomes many limitations of
previous proposals \citep{kolchinskyNovelApproachPartial2022}.

The \emph{information bottleneck} (IB) \citep{tishbyInformationBottleneckMethod1999,hu2024survey}
is a method for extracting compressed information from one random
variable $X$ that optimally predicts another target random variable
$Y$. For instance, in the neuroscience example with stimulus $Y$
and brain activity $X$, the IB method could be used to quantify how
well the stimulus can be predicted using only one bit of information
about brain activity. The overall tradeoff between the prediction of $Y$
and compression of $X$ is captured by the so-called \emph{IB curve}.
The IB method has been employed in various domains, including neuroscience
\citep{palmer2015predictive}, biology \citep{wangFutureInformationBottleneck2019},
and cognitive science \citep{zaslavskyEfficientCompressionColor2018}.
In recent times, it has become particularly popular in machine learning
applications \citep{alemi2016deep,kolchinsky2019nonlinear,fischer2020conditional,goldfeld2020information,hu2024survey}. 

In this paper, we demonstrate a formal connection between PID and 
IB. We focus in particular on the relationship between the IB and PID
redundancy, leaving the connection to other PID measures (such as
synergy) for future work. To begin, we show that Blackwell redundancy
can be formulated as an information-theoretic constrained optimization
problem. This optimization problem extracts information from the sources
that best predict the target, under the constraint that the solution
does not reveal which source provided the information. We then define
a generalized measure of Blackwell redundancy by relaxing the constraint.
Specifically, we ask how much predictive information can be extracted
from the sources without revealing more than a certain number of bits
about the identity of the source. Our generalization leads to an IB-type
tradeoff between the prediction of the target (generalized redundancy)
and compression (leakage of information about the identity of the
source). We refer to the resulting optimization problem as the 
\emph{redundancy bottleneck} (RB) and to the manifold of optimal solutions
at different points on the prediction/compression tradeoff as the
\emph{RB curve}. We also show that the RB prediction and compression
terms can be decomposed into contributions from individual sources, giving
rise to an individual RB curve for each source.

Besides the intrinsic theoretical interest of unifying PID and the IB,
our approach brings about several practical advantages. In particular,
the RB curve offers a fine-grained analysis on PID redundancy, showing
how redundant information emerges at various scales and across different
sources. This fine-grained analysis can be used to uncover sets of
redundant sources without performing intractable combinatorial optimization.
Our approach also has numerical advantages. The original formulation
of Blackwell redundancy was based on a difficult optimization problem
that becomes infeasible for larger systems. By reformulating Blackwell
redundancy as an IB-type problem, we are able to solve it efficiently
using an iterative algorithm, even for larger systems (code available
at \url{https://github.com/artemyk/pid-as-ib}, accessed 25-06-2024). Finally, the RB has some attractive formal properties. For instance, unlike
the original Blackwell redundancy, the RB curve is continuous in the
underlying probability distributions.

This paper is organized as follows. In the next section, we provide the
background on the IB, PID, and Blackwell redundancy. In Section \ref{sec:Redundancy-bottleneck},
we introduce the RB, illustrate it with several examples, and discuss
its formal properties. In Section \ref{sec:algo}, we introduce an
iterative algorithm to solve the RB optimization problem. We discuss the implications
and possible future directions in Section \ref{sec:Discussion}. All
proofs are found in the Appendix.

\section{Background}

\label{sec:background}

We begin by providing relevant background on the information bottleneck,
partial information decomposition, and Blackwell redundancy.

\subsection{Information Bottleneck (IB)}

The information bottleneck (IB) method provides a way to extract information
that is present in one random variable $X$ that is relevant for predicting
another target random variable $Y$ \citep{tishbyInformationBottleneckMethod1999,ahlswedeSourceCodingSide1975,witsenhausen_conditional_1975}.
To do so, the IB posits a “bottleneck variable'' $Q$ that obeys the
Markov condition $Q-X-Y$. This Markov condition guarantees that $Q$
does not contain any information about $Y$ that is not found in $X$.
The quality of any particular choice of bottleneck variable $Q$ is
quantified via two mutual information terms: $I(X;Q)$, which decreases
when $Q$ provides a more compressed representation of $X$, and $I(Y;Q)$,
which increases when $Q$ allows a better prediction of $Y$. The
IB method selects $Q$ to maximize prediction given a constraint on
compression \citep{witsenhausen_conditional_1975,ahlswedeSourceCodingSide1975,goos_information_2003}:
\begin{equation}
I_{\text{IB}}(R)=\max_{Q:Q-X-Y}I(Y;Q)\where I(X;Q)\le R.\label{eq:ibproblem}
\end{equation}
The values of $I_{\mathrm{IB}}(R)$ for different $R$ specify the
\emph{IB curve}, which encodes the overall tradeoff between prediction
and compression.

In practice, the IB curve is usually explored by considering the Lagrangian
relaxation of the constrained optimization problem (\ref{eq:ibproblem}):
\begin{equation}
F_{\text{IB}}(\ibL):=\max_{Q}I(Y;Q)-\frac{1}{\ibL}I(X;Q)\label{eq:ibbeta}
\end{equation}
Here, $\ibL\ge0$ is a parameter that controls the tradeoff between
compression cost (favored for $\ibL\to0$) and prediction benefit
(favored for $\ibL\to\infty$). The advantage of the Lagrangian formulation
is that it avoids the non-linear constraint in \eqname(\ref{eq:ibproblem}).
If the IB curve is strictly concave, then the two formulations (\ref{eq:ibproblem})
and (\ref{eq:ibbeta}) are equivalent, meaning that there is a one-to-one
map between the solutions of both problems \citep{kolchinskyCaveatsInformationBottleneck2019}.
When the IB curve is not strictly concave, a modified objective such
as the ``squared Lagrangian'' or ``exponential Lagrangian'' should
be used instead; see Refs. \citep{kolchinskyCaveatsInformationBottleneck2019,rodriguez2020convex,benger2023cardinality}
for more details.

Since the original proposal, many reformulations, generalizations,
and variants of the IB have been developed \citep{hu2024survey}.
Notable examples include the ``conditional entropy bottleneck''
(CEB) \citep{fischer2020conditional,geiger2020comparison}, the ``multi-view
IB'' \citep{federici2020learning}, the ``distributed IB'' \citep{murphy2024machine},
as well as a large family of objectives called the ``multivariate
IB'' \citep{slonimMultivariateInformationBottleneck2006}. All of
these approaches consider some tradeoff between two information-theoretic
terms: one that quantifies the prediction of target information that
should be maximized and one that quantifies the compression of unwanted
information that should be minimized. We refer to an optimization
that involves a tradeoff between information-theoretic prediction
and compression terms as an \emph{IB-type problem}.

\subsection{Partial Information Decomposition}

The PID considers how information about a target random variable $Y$
is distributed across a set of source random variables $X_{1},\dots,X_{n}$.
One of the main goals of the PID is to quantify redundancy, the amount
of shared information that is found in each of the individual sources.
The notion of redundancy in PID was inspired by an analogy between
sets and information that has re-appeared in various
forms throughout the history of information theory \citep{shannon_lattice_1953,mcgill1954multivariate,rezaIntroductionInformationTheory1961,ting1962amount,han1975linear,yeung1991new,bell2003co}.
Specifically, if the amount of information provided by each source
is conceptualized as the size of a set, then the redundancy is conceptualized
as the size of the intersection of those sets \citep{williams2010nonnegative,williams_information_2011,kolchinskyNovelApproachPartial2022}.
Until recently, however, this analogy was treated mostly as an informal
source of intuition, rather than a formal methodology.

In a recent paper \citep{kolchinskyNovelApproachPartial2022}, we
demonstrated that the terms of PID can be defined by formalizing
this analogy to set theory. Recall that, in set theory, the intersection
of sets $A_{1},\dots,A_{n}$ is defined as the largest set $B$ that
is contained in each set $A_{s}$ for $s\in\{1...n\}$. Thus, the
size of the intersection of finite sets $A_{1},\dots,A_{n}$ is
\[
\left|\bigcap_{\vvss=1}^{n}A_{\vvss}\right|=\max_{B}\vert B\vert\where B\subseteq A_{\vvss}\quad\forall s\in\{1\dots n\}.
\]
We showed that PID redundancy can be defined in a similar way: the
redundancy between sources $X_{1},\dots,X_{n}$ is the maximum mutual
information in any random variable $Q$ that is less informative about
the target $Y$ than each individual source \citep{kolchinskyNovelApproachPartial2022}:
\begin{equation}
\RR^{\sqsubseteq}:=\max_{Q}I(Q;Y)\where Q\sqsubseteq X_{\vvss}\quad\forall s\in\{1\dots n\}.\label{eq:red}
\end{equation}
The notation $Q\sqsubseteq X_{\vvss}$ indicates that $Q$ is ``less informative''
about the target than $X_{\vvss}$, given some pre-specified 
ordering relation $\sqsubseteq$. 
The choice of the ordering relation completely determines the
resulting redundancy measure $\RR^{\sqsubseteq}$. We discuss possible choices in the following subsection.

We used a similar approach to define ``union information'', which
in turn leads to a principled measure of synergy \citep{kolchinskyNovelApproachPartial2022}.
Note that union information and redundancy are related algebraically
but not numerically; in particular, unlike in set theory, the principle
of inclusion--exclusion does not always hold.

As mentioned above, here, we focus entirely on redundancy and leave
the exploration of connections between IB and union information/synergy
for future work.

\subsection{Blackwell Redundancy}

\label{subsec:Blackwell-redundancy}

\global\long\def\chB{\kappa_{B\vert Y}}%
\global\long\def\chC{\kappa_{C\vert Y}}%
Our definition of PID redundancy (\ref{eq:red}) depends on the definition
of the ``less informative'' relation $\sqsubseteq$. Although there
are many relations that can be considered \citep{gomes2023orders,griffith2014quantifying,griffith2014intersection,griffith_quantifying_2015,shannon_lattice_1953},
arguably the most natural choice is the \emph{Blackwell order}. 

The Blackwell order is a preorder relation over ``channels'', that is conditional
distributions with full support. A channel $\chB$ is said to be less
informative than $\chC$ in the sense of the Blackwell order if there
exists some other channel $\kappa_{B\vert C}$ such that
\begin{align}
\chB & =\kappa_{B\vert C}\circ\chC.\label{eq:garbl}
\end{align}
Throughout, we use the notation $\circ$ to indicate the composition of
channels, as defined via matrix multiplication. For instance, $\chB=\kappa_{B\vert C}\circ\chC$
is equivalent to the statement $\chB(b\vert y)=\sum_{c}\kappa_{B\vert C}(b\vert c)\chC(c\vert y)$
for all $b$ and $y$. \eqname(\ref{eq:garbl}) implies that $\chB$
is less informative than $\chC$ if $\chB$ can be produced by downstream
stochastic processing of the output of channel $\chC$. We use the
notation
\begin{equation}
\chB\preceq\chC,\label{eq:blackwell}
\end{equation}
to indicate that $\chB$ is less Blackwell-informative than $\chC$.
The Blackwell order can also be defined over random variables rather
than channels. Given a target random variable $Y$ with full support,
random variable $B$ is said to be less Blackwell-informative than
random variable $C$, written as
\begin{equation}
B\preceq_{Y}C,\label{eq:blackwellrv}
\end{equation}
when their corresponding conditional distributions obey the Blackwell
relation, $p_{B\vert Y}\preceq p_{C\vert Y}$ \citep{bertschinger2014blackwell}.
It is not hard to verify that any random variable $B$ that is independent of $Y$ is lowest under the Blackwell order, obeying $B\preceq_{Y}C$ for all $C$.

The Blackwell order plays a fundamental role in statistics, and it
has an important operational characterization in decision theory \citep{blackwell_equivalent_1953,bertschinger2014blackwell,rauh_coarse-graining_2017}.
Specifically, $p_{B\vert Y}\preceq p_{C\vert Y}$ if and only if access
to channel $p_{C\vert Y}$ is better for every decision problem than
access to channel $p_{B\vert Y}$. See Refs. \citep{bertschinger2014quantifying,kolchinskyNovelApproachPartial2022}
for details of this operational characterization and Refs. \citep{bertschinger2014blackwell,bertschinger2014quantifying,rauh2017extractable,kolchinskyNovelApproachPartial2022,venkatesh2022partial,mages2024non}
for more discussion of the relation between the Blackwell order and the PID. 

Combining the Blackwell order (\ref{eq:blackwellrv}) with \eqname(\ref{eq:red})
gives rise to \emph{Blackwell redundancy} \citep{kolchinskyNovelApproachPartial2022}.
Blackwell redundancy, indicated here as $\RR$, is the maximal mutual
information in any random variable that is less Blackwell-informative
than each of the sources:
\begin{equation}
I_{\cap}:=\max_{Q}\,I(Q;Y)\where Q\preceq_{Y}X_{\vvss}\;\forall\vvss.\label{eq:opt1}
\end{equation}
The optimization is always well defined because the feasible set is not empty, given that any random variable $Q$ that is independent of $Y$ satisfies the constraints. (Note also that, for continuous-valued or countably infinite sources, $\max$ may need to be replaced by a $\sup$; see also Appendix~\ref{app:1}.)

$I_{\cap}$ has many attractive features as a measure of PID redundancy,
and it overcomes several problems with previous approaches \citep{kolchinskyNovelApproachPartial2022}.
For instance, it can be defined for any number of sources, it uniquely
satisfies a natural set of PID axioms, and it has fundamental statistical
and operational interpretations. Statistically, it is the maximum information
transmitted across any channel that can be produced by downstream
processing of any one of the sources. Operationally, it is the maximum
information that any random variable can have about $Y$ without being able
to perform better on any decision problem than any one of the sources.

As we showed \citep{kolchinskyNovelApproachPartial2022}, the optimization
problem (\ref{eq:opt1}) can be formulated as the maximization of
a convex function subject to a set of linear constraints. For a finite-dimensional
system, the feasible set is a finite-dimensional polytope, and the
maximum will lie on one of its extreme points; therefore, the optimization
can be solved exactly by enumerating the vertices of the feasible
set and choosing the best one \citep{kolchinskyNovelApproachPartial2022}.
However, this approach is limited to small systems, because the number
of vertices of the feasible set can grow exponentially. 

Finally, it may be argued that Blackwell
redundancy is actually a measure of redundancy in the channels $p_{X_{1}\vert Y},\dots,p_{X_{n}\vert Y}$,
rather than in the random variables $X_{1},\dots,X_{n}$. This is
because the joint distribution over $(Y,X_{1},\dots,X_{n})$ is never
explicitly invoked in the definition of $I_{\cap}$; in fact, any
joint distribution is permitted as long as it is compatible with the
correct marginals. (The same property holds for several other redundancy
measures \citep[Table 1, ][]{kolchinskyNovelApproachPartial2022},
and Ref.~\citep{bertschinger2014quantifying} even suggested this
property as a requirement for any valid measure of PID redundancy.)  
In some cases, the joint distribution may not even exist, for instance
when different sources represent mutually exclusive conditions. To
use a neuroscience example, imagine that $p_{X_{1}\vert Y}$ and $p_{X_{2}\vert Y}$
represent the activity of some brain region $X$ in response to stimulus
$Y$, measured either in younger ($p_{X_{1}\vert Y}$) or older ($p_{X_{2}\vert Y}$)
subjects. Even though there is no joint distribution over $(Y,X_1,X_2)$ in this case, redundancy
is still meaningful as the information about the stimulus
that can be extracted from the brain activity of either age group.
In the rest of this paper, we generally work within the channel-based
interpretation of Blackwell redundancy.

\section{Redundancy Bottleneck}

\label{sec:Redundancy-bottleneck}

In this section, we introduce the redundancy bottleneck (RB) and illustrate
it with examples. Generally, we assume that we are provided
with the marginal distribution $p_{Y}$ of the target random variable
$Y$, as well as $n$ source channels $p_{X_{1}\vert Y},\dots,p_{X_{n}\vert Y}$.
Without loss of generality, we assume that $p_{Y}$ has full support.
We use calligraphic letters (like $\mathcal{Y}$ and $\mathcal{X}_{\vvss}$)
to indicate the set of outcomes of random variables (like $Y$ and
$X_{\vvss}$). For simplicity, we use notation appropriate
for discrete-valued variables, such as in \eqname(\ref{eq:garbl}),
though most of our results also apply to continuous-valued variables.

\subsection{Reformulation of Blackwell Redundancy}

\global\long\def\redX{Z}%
\global\long\def\redx{z}%

We first reformulate Blackwell redundancy (\ref{eq:opt1}) in terms
of a different optimization problem. Our reformulation will make use
of the random variable $Y$, along with two additional random variables,
$S$ and $\redX$. The outcomes of $S$ are the indexes of the different
sources, $\mathcal{S}=\{1,\dots,n\}$. The set of outcomes of $\redX$
is the union of the outcomes of the individual sources, $\mathcal{\redX}=\bigcup_{\vvss=1}^{n}\mathcal{X}_{\vvss}$.
For example, if there are two sources with outcomes $\mathcal{X}_{1}=\{0,1\}$
and $\mathcal{X}_{2}=\{0,1,2\}$, then $\mathcal{S}=\{1,2\}$ and
$\mathcal{Z}=\{0,1\}\cup\{0,1,2\}=\{0,1,2\}$. The joint probability
distribution over $(Y,S,\redX)$ is defined as
\begin{equation}
p_{YS\redX}(y,s,\redx)=\begin{cases}
p_{Y}(y)\sdist(s)p_{X_{s}\vert Y}(\redx\vert y) & \text{if }\redx\in\mathcal{X}_{s}\\
0 & \text{otherwise}
\end{cases}\label{eq:pdef}
\end{equation}
In other words, $y$ is drawn from the marginal $p_{Y}$, the source $s$
is then drawn independently from the distribution $\sdist$, and finally 
$\redx$ is drawn from the channel $p_{X_{s}\vert Y}(\redx\vert y)$
corresponding to source $s$. In this way, the channels corresponding
to the $n$ sources ($p_{X_{1}\vert Y},\dots,p_{X_{n}\vert Y}$) are
combined into a single conditional distribution $p_{\redX\vert SY}$. 

We treat the distribution $\sdist$ as an arbitrary fixed parameter,
and except where otherwise noted, we make no assumptions about this
distribution except that it has full support. As we will
see, different choices of $\sdist$ cause the different sources
to be weighed differently in the computation of the RB. We return
to the question of how to determine this distribution below. 

Note that, under the distribution defined in \eqname(\ref{eq:pdef}),
$Y$ and $S$ are independent, so
\begin{equation}
I(Y;S)=0.\label{eq:YSind}
\end{equation}
Actually, many of our results can be generalized to the case where
there are correlations between $S$ and $Y$. We leave exploration
of this generalization for future work.

In addition to $Y$, $\redX$, and $S$, we introduce another random
variable $Q$. This random variable obeys the Markov condition $Q-(\redX,S)-Y$,
which ensures that $Q$ does not contain any information about $Y$
that is not contained in the joint outcome of $\redX$ and $S$. The
full joint distribution over $(Y,S,\redX,Q)$ is 
\begin{equation}
p_{YS\redX Q}(y,s,\redx,q)=p_{YS\redX}(y,s,\redx)p_{Q\vert S\redX}(q\vert s,\redx).\label{eq:jointd}
\end{equation}
We sometimes refer to $Q$ as the ``bottleneck'' random variable.

The set of joint outcomes of $(S,Z)$ with non-zero probability
is the disjoint union of the outcomes of the individual sources. For
instance, in the example above with $\mathcal{X}_{1}=\{0,1\}$ and
$\mathcal{X}_{2}=\{0,1,2\}$, the set of joint outcomes of $(S,Z)$
with non-zero probability is $\{(1,0),(1,1),(2,0),(2,1),(2,2)\}$.
Because $Q$ depends jointly on $S$ and $\redX$, our results do
not depend on the precise labeling of the source outcomes, e.g.,
they are the same if $\mathcal{X}_{2}=\{0,1,2\}$ is relabeled
as $\mathcal{X}_{2}=\{2,3,4\}$.

Our first result shows that Blackwell redundancy can be equivalently
expressed as a constrained optimization problem. Here, the optimization
is over bottleneck random variables $Q$, i.e., over conditional distributions
$p_{Q\vert S\redX}$ in \eqname(\ref{eq:jointd}).
\begin{thm}
\label{thm:1}Blackwell redundancy (\ref{eq:opt1}) can be expressed
as
\begin{align}
I_{\cap}=\max_{Q:Q-(\redX,S)-Y}I(Q;Y\vert S)\where I(Q;S\vert Y)=0.
\label{eq:thm1}
\end{align}
\end{thm}
Importantly, Theorem~\ref{thm:1} does not depend on the choice of
the distribution $\sdist$, as long as it has full support.

In Theorem~\ref{thm:1}, the Blackwell order constraint in \eqname(\ref{eq:opt1})
has been replaced by an information-theoretic constraint $I(Q;S\vert Y)=0$,
which states that $Q$ does not provide any information about the
identity of source $S$, additional to that already provided by the
target $Y$. The objective $I(Q;Y)$ has been replaced by the conditional
mutual information $I(Q;Y\vert S)$. Actually, the objective can be
equivalently written in either form, since $I(Q;Y\vert S)=I(Q;Y)$
given our assumptions (see the proof of Theorem~\ref{thm:1} in the Appendix).
However, the conditional mutual information form will be useful for
further generalization and decomposition, as discussed in the next
sections.

\subsection{Redundancy Bottleneck}

To relate Blackwell redundancy to the IB, we relax the constraint in Theorem~\ref{thm:1}
by allowing the leakage of $R$ bits of conditional information about
the source $S$. This defines the \emph{redundancy bottleneck} (RB)
at compression rate $\rrate$:
\begin{equation}
\RBR:=\quad\max_{\mathclap{Q:Q-(\redX,S)-Y}}\quad I(Q;Y\vert S)\where I(Q;S\vert Y)\le\rrate.\label{eq:opt2-1}
\end{equation}
We note that, for $R>0$, the value of $\RBR$ does depend on the choice
of the source distribution $\sdist$.

\eqname(\ref{eq:opt2-1}) is an IB-type problem that involves a tradeoff
between prediction $I(Q;Y\vert S)$ and compression $I(Q;S\vert Y)$.
The prediction term $I(Q;Y\vert S)$ quantifies the generalized Blackwell
redundancy encoded in the bottleneck variable $Q$. The compression
term $I(Q;S\vert Y)$ quantifies the amount of conditional information
that the bottleneck variable leaks about the identity of the source.
The set of optimal values of $(I(Q;S\vert Y),I(Q;Y\vert S))$ defines
the \emph{redundancy bottleneck curve} (RB curve) that encodes the
overall tradeoff between prediction and compression. 

We prove a few useful facts about the RB, starting from monotonicity
and concavity.
\begin{thm}
\label{thm:conc}$\RBR$ is non-decreasing and concave as a function
of $\rrate$.
\end{thm}
Since $\RBR$ is non-decreasing in $R$, the lowest RB value is achieved
in the $R=0$ regime, when it equals the Blackwell redundancy (Theorem~\ref{thm:1}):
\begin{equation}
\RBR\ge\RB 0=\RR.\label{eq:lb}
\end{equation}
The largest value is achieved as $\rrate\to\infty$, when the compression
constraint vanishes. It can be shown that $I(Q;Y\vert S)\le I(\redX;Y\vert S)=I(Y;Z,S)$
using the Markov condition $Q-(Z,S)-Y$ and the data-processing inequality
(see the next subsection). This upper bound is achieved by the
bottleneck variable $Q=\redX$. Combining implies
\begin{equation}
\RBR\le I(\redX;Y\vert S)=\sum_{s}\sdist(s)I(X_{\vvss};Y),\label{eq:up}
\end{equation}
where we used the form of the distribution $p_{YS\redX}$ in \eqname(\ref{eq:pdef})
to arrive at the last expression. The range of necessary compression
rates can be restricted as $0\le R\le I(Z;S\vert Y)$.

 Next, we show that, for finite-dimensional sources, it suffices to
consider finite-dimensional $Q$. Thus, for finite-dimensional
sources, the RB problem (\ref{eq:opt2-1}) involves the maximization
of a continuous objective over a compact domain, so the maximum is
always achieved by some $Q$. (Conversely, in the more general case of infinite-dimensional sources, it may be necessary to replace $\max$ with $\sup$ in \eqname\eqref{eq:opt2-1}; see Appendix~\ref{app:1}.)
\begin{thm}
\label{thm:card}For the optimization problem (\ref{eq:opt2-1}),
it suffices to consider $Q$ of cardinality $\left|\mathcal{Q}\right|\le\sum_{\vvss}\left|\mathcal{X}_{\vvss}\right|+1$.
\end{thm}
Interestingly, the cardinality bound for the RB is the same as for the IB
if we take $X=(\redX,S)$ in \eqname(\ref{eq:ibproblem}) \citep{witsenhausen_conditional_1975,benger2023cardinality}.
It is larger than the cardinality required for Blackwell redundancy
(\ref{eq:opt1}), where $\vert\mathcal{Q}\vert\le(\sum_{\vvss}\left|\mathcal{\mathcal{X}}_{\vvss}\right|)-n+1$
suffices \citep{kolchinskyNovelApproachPartial2022}.

The Lagrangian relaxation of the constrained RB problem (\ref{eq:opt2-1})
is given by
\begin{equation}
F_{\text{RB}}(\cebL)=\max_{Q:Q-(\redX,S)-Y}I(Q;Y\vert S)-\frac{1}{\beta}I(Q;S\vert Y).\label{eq:opt3}
\end{equation}
The parameter $\cebL$ controls the tradeoff between prediction and
compression. The $\cebL\to0$ limit corresponds to the $R=0$ regime,
in which case, Blackwell redundancy is recovered, while the $\cebL\to\infty$
limit corresponds to the $R=\infty$ regime, when the compression
constraint is removed. The RB Lagrangian (\ref{eq:opt3}) is often
simpler to optimize than the constrained optimization (\ref{eq:opt2-1}).
Moreover, when the RB curve $\RBR$ is strictly concave, there is
a one-to-one relationship between the solutions to the two optimization
problems (\ref{eq:opt2-1}) and (\ref{eq:opt3}). However, when the
RB curve is not strictly concave, there is no one-to-one relationship
and the usual Lagrangian formulation is insufficient. This can be
addressed by optimizing a modified objective that combines prediction
and compression in a nonlinear fashion, such as the ``exponential
Lagrangian'' \citep{rodriguez2020convex}: 
\begin{equation}
F_{\text{RB}}^{\exp}(\cebL)=\max_{Q:Q-(\redX,S)-Y}I(Q;Y\vert S)-\frac{1}{\beta}e^{I(Q;S\vert Y)}.\label{eq:RBlagrangianExp}
\end{equation}
(See an analogous analysis for IB in Refs.~\citep{kolchinskyCaveatsInformationBottleneck2019,rodriguez2020convex}.)

\subsection{Contributions From Different Sources}

Both the RB prediction and compression terms can be decomposed into
contributions from different sources, leading to an individual RB
curve for each source. As we show in the examples below, this decomposition
can be used to identify groups of redundant sources without having
to perform intractable combinatorial optimization.

\begin{figure*}
\includegraphics[width=1\textwidth]{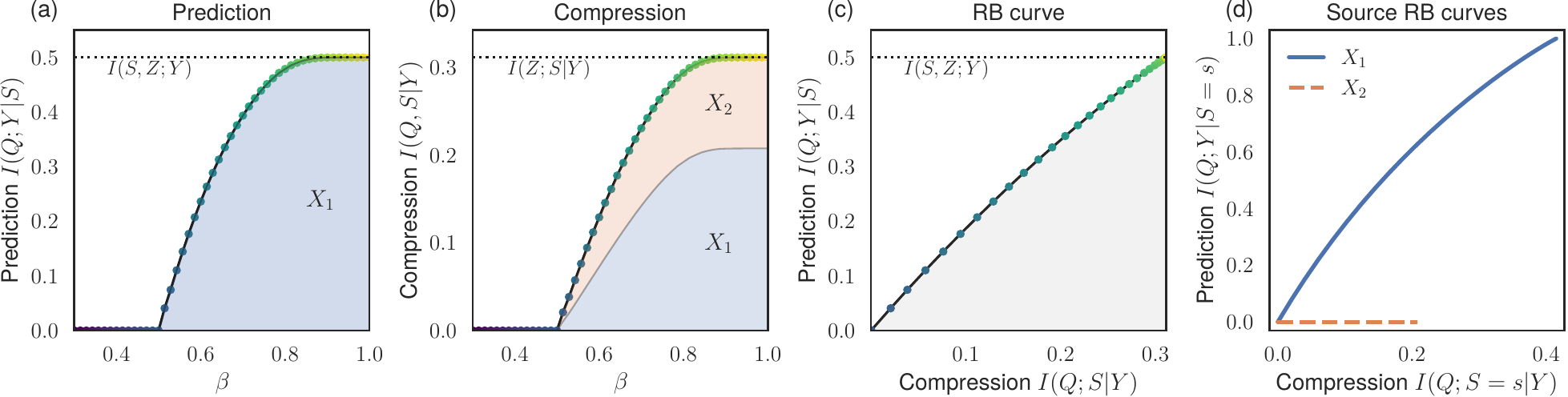}

\caption{\protect\label{fig:unq}RB analysis for the UNIQUE gate
(Example 1). \textbf{(a)} Prediction values found by optimizing the
RB Lagrangian (\ref{eq:opt3}) at different $\beta$. Colored regions
indicate contributions from different sources, $\protect\sdist(s)I(Q;Y\vert S=s)$
from \eqname(\ref{eq:decomp0-1}). For this system, only source $X_{1}$
contributes to the prediction. \textbf{(b)} Compression costs found by
optimizing the RB Lagrangian at different $\beta$. Colored regions
indicate contributions from different sources, $\protect\sdist(s)I(Q;S=s\vert Y)$
from \eqname(\ref{eq:decompC}). \textbf{(c)} The RB curve shows the
tradeoff between optimal compression and the prediction values; the marker
colors correspond to the $\beta$ values as in (a) and (b). All bottleneck
variables $Q$ must fall within the accessible grey region. \textbf{(d)
}RB curves for individual sources.}
\end{figure*}

Let $Q$ be an optimal bottleneck variable at rate $R$, so that $\RBR=I(Q;Y\vert S)$
and $I(Q;S\vert Y)\le R$. Then, the RB prediction term can be expressed
as the weighted average of the prediction contributions from individual
sources:
\begin{align}
\RBR=I(Q;Y\vert S) & =\sum_{\vvss}\sdist(s)I(Q;Y\vert S=s).\label{eq:decomp0-1}
\end{align}
Here, we introduce the specific conditional mutual information:
\begin{equation}
I(Q;Y\vert S=s):=D(p_{Q\vert Y,S=\vvss}\Vert p_{Q\vert S=\vvss}),\label{eq:specificmi}
\end{equation}
where $D(\cdot\Vert\cdot)$ is the Kullback--Leibler (KL) divergence.
To build intuitions about this decomposition, we may use the Markov condition
$Q-(\redX,S)-Y$ to express the conditional distributions in \eqname(\ref{eq:specificmi})
as compositions of channels:
\begin{align*}
p_{Q\vert Y,S=\vvss} & =p_{Q\vert\redX,S=s}\circ p_{\redX\vert Y,S=\vvss}\\
p_{Q\vert S=\vvss} & =p_{Q\vert\redX,S=s}\circ p_{\redX\vert S=\vvss}
\end{align*}
Using the data-processing inequality for the KL divergence and \eqname(\ref{eq:pdef}),
we can then write
\[
I(Q;Y\vert S=s)\le D(p_{\redX\vert Y,S=\vvss}\Vert p_{\redX\vert S=\vvss})=D(p_{X_{s}\vert Y}\Vert p_{X_{s}}).
\]
The last term is simply the mutual information $I(Y;X_{s})$ between
the target and source $s$. Thus, the prediction contribution from
source $s$ is bounded between 0 and the mutual information provided
by that source:
\begin{equation}
0\le I(Q;Y\vert S=s)\le I(Y;X_{s}).\label{eq:bnds0}
\end{equation}
The difference between the mutual information and the actual prediction
contribution:
\[
I(Y;X_{s})-I(Q;Y\vert S=s)\ge0,
\]
quantifies the unique information in source $s$. The upper bound
in \eqname(\ref{eq:bnds0}) is achieved in the $R\to\infty$ limit by
$Q=Z$, leading to \eqname(\ref{eq:up}). Conversely, for $R=0$, $p_{Q\vert Y,S=\vvss}=p_{Q\vert Y}$
(from $I(Q;S\vert Y)=0$) and $p_{Q\vert S=\vvss}=p_{Q}$ (from \eqname(\ref{eq:YSind})),
so 
\[
I(Q;Y\vert S=s)=I(Q;Y)=\RB 0=\RR.
\]
Thus, when $R=0$, the prediction contribution from each source is
the same, and it is equal to the Blackwell redundancy. 

The RB compression cost can also be decomposed into contributions
from individual sources:
\begin{align}
I(Q;S\vert Y) & =\sum_{\vvss}\sdist(s)I(Q;S=s\vert Y).\label{eq:decompC}
\end{align}
Here, we introduce the specific conditional mutual information:
\begin{equation}
I(Q;S=s\vert Y):=D(p_{Q\vert Y,S=\vvss}\Vert p_{Q\vert Y}).\label{eq:specificmi-1}
\end{equation}

The source compression terms can be related to so-called \emph{deficiency},
a quantitative generalization of the Blackwell order. Although various
versions of deficiency can be defined \citep{le1964sufficiency,raginsky_shannon_2011,banerjee_unique_2018},
here we consider the ``weighted deficiency'' induced by the KL divergence.
For any two channels $p_{B\vert Y}$ and $p_{C\vert Y}$, it is defined
as
\begin{align}
\delta_{D}(p_{C\vert Y},p_{B\vert Y}) & :=\min_{\kappa_{B\vert C}}D(\kappa_{B\vert C}\circ p_{C\vert Y}\Vert p_{B\vert Y}).\label{eq:defdef}
\end{align}
This measure quantifies the degree to which two channels violate the
Blackwell order, vanishing when $\kappa_{B\vert Y}\preceq\kappa_{C\vert Y}$.
To relate the source compression terms (\ref{eq:decompC}) to deficiency,
observe that $p_{Q\vert Y,S=\vvss}=p_{Q\vert\redX,S=s}\circ p_{\redX\vert Y,S=\vvss}$
and that $p_{\redX\vert Y,S=\vvss}=p_{X_{s}\vert Y}$. Given \eqname(\ref{eq:specificmi-1}),
we then have
\begin{equation}
I(Q;S=\vvss\vert Y)\ge\delta_{D}(p_{X_{s}\vert Y},p_{Q\vert Y}).\label{eq:defbnd}
\end{equation}
Thus, each source compression term is lower bounded by the deficiency
between the source channel $p_{X_{s}\vert Y}$ and the bottleneck
channel $p_{Q\vert Y}$. Furthermore, the compression constraint in
the RB optimization problem (\ref{eq:opt2-1}) sets an upper bound
on the deficiency of $p_{Q\vert Y}$ averaged across all sources.

Interestingly, several recent papers have studied the relationship
between deficiency and PID redundancy in the restricted case of two
sources \citep{rauh_coarse-graining_2017,banerjee_unique_2018,banerjeeVariationalDeficiencyBottleneck2020,venkatesh2022partial,venkateshCapturingInterpretingUnique2023}.
To our knowledge, we provide the first link between deficiency and
redundancy for the general case of multiple sources. Note also that
previous work considered a slightly different definition of deficiency
where the arguments of the KL divergence are reversed. Our definition
of deficiency is arguably more natural, since it is more natural to
minimize the KL divergence 
over a convex set with respect to the first argument \citep{csiszarInformationProjectionsRevisited2003}.

Finally, observe that, in both decompositions (\ref{eq:decomp0-1})
and (\ref{eq:decompC}), the source contributions are weighted by
the distribution $\sdist(s)$. Thus, the distribution $\sdist$ determines
how different sources play into the tradeoff between prediction and
compression. In many cases, $\sdist$ can be chosen as
the uniform distribution. However, other choices of $\sdist$ may
be more natural in other situations. For example, in a neuroscience
context where different sources correspond to different brain regions,
$\sdist(s)$ could represent the proportion of metabolic cost or neural
volume assigned to region $s$. Alternatively, when different
sources represent mutually exclusive conditions, as in the age group
example mentioned at the end of Section \ref{sec:background}, $\sdist(s)$
might represent the frequency of condition $s$ found in the data.
Finally, it may be possible to set $\sdist$ in an ``adversarial''
manner so as to maximize the resulting value of $\RB R$ in \eqname(\ref{eq:opt2-1}).
We leave the exploration of this adversarial approach for future
work.

\subsection{Examples}

We illustrate our approach using a few examples. For simplicity, in
all examples, we use a uniform distribution over the sources, $\sdist(s)=1/n$.
The numerical results are calculated using the iterative algorithm described
in the next section.

\vspace{5pt}
\noindent \textbf{Example 1.} We begin by considering a very simple system, called the ``UNIQUE gate''
in the PID literature. Here, the target $Y$ is binary and uniformly-distributed,
$p_{Y}(y)=1/2$ for $y\in\{0,1\}$. There are two binary-valued sources,
$X_{1}$ and $X_{2}$, where the first source is a copy of the target,
$p_{X_{1}\vert Y}(x_{1}\vert y)=\delta_{x_{1},y}$, while the second
source is an independent and uniformly-distributed bit, $p_{X_{2}\vert Y}(x_{1}\vert y)=1/2$.
Thus, source $X_{1}$ provides 1 bit of information about the target,
while $X_{2}$ provides none. The Blackwell redundancy is $\RR=0$
\citep{kolchinskyNovelApproachPartial2022}, because it is impossible
to extract any information from the sources without revealing that
this information came from $X_{1}$.

\begin{figure*}
\includegraphics[width=1\textwidth]{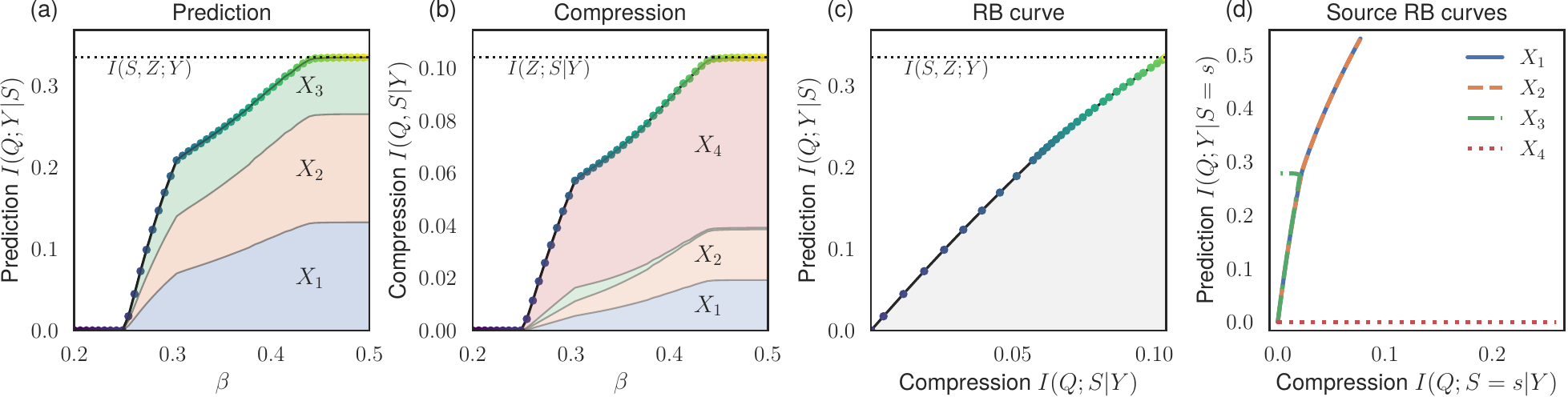}

\caption{\protect\label{fig:4src}RB analysis for the system with
4 binary symmetric channels (Example 3). \textbf{(a)} and \textbf{(b)}
Prediction and compression values found by optimizing the RB Lagrangian
(\ref{eq:opt3}) at different $\beta$. Contributions from individual
sources are shown as shaded regions. \textbf{(c)} The RB curve shows the tradeoff between optimal compression and
prediction values; marker colors correspond to the $\beta$ values as
in (a) and (b). \textbf{(d)} RB curves for individual sources.}
\end{figure*}

We performed RB analysis by optimizing the RB Lagrangian $F_{\text{RB}}(\cebL)$
(\ref{eq:opt3}) at different $\beta$. Figure~\ref{fig:unq}(a)
and (b) show the prediction $I(Q;Y\vert S)$ and compression $I(Q;S\vert Y)$
values for the optimal bottleneck variables $Q$. At small $\beta$, the
prediction converges to the Blackwell redundancy, $I(Q;Y\vert S)=\RR=0$,
and there is complete loss of information about source identity, $I(Q;S\vert Y)=0$.
At larger $\beta$, the prediction approaches the maximum $I(Q;Y\vert S)=0.5\times I(X_{1};Y)=0.5\text{ bit}$,
and compression approaches $I(Q;S\vert Y)=I(Z;S\vert Y)\approx0.311\text{ bit}$.
Figure~\ref{fig:unq}(c) shows the RB curve, illustrating the overall
tradeoff between prediction and compression. 

In the shaded regions of Figure~\ref{fig:unq}(a) and (b), we show
the additive contributions to the prediction and compression terms from
the individual sources, $\sdist(s)I(Q;Y\vert S=s)$ from \eqname(\ref{eq:decomp0-1})
and $\sdist(s)I(Q;S=s\vert Y)$ from \eqname(\ref{eq:decompC}), respectively.
We also show the resulting RB curves for individual sources in Figure~\ref{fig:unq}(d).
As expected, only source $X_{1}$ contributes to the prediction at any
level of compression.

To summarize, if some information about the identity of the source
can be leaked (non-zero compression cost), then improved prediction
of the target is possible. At the maximum needed compression cost
of $0.311$, it is possible to extract 1 bit of predictive information
from $X_{1}$ and 0 bits from $X_{2}$, leading to an average of $0.5\text{ bits}$
of prediction.

%

\vspace{5pt}
\noindent \textbf{Example 2.} 
We now consider the ``AND gate'', another well-known system from
the PID literature. There are two independent and uniformly distributed
binary sources, $X_{1}$ and $X_{2}$. The target $Y$ is also binary-valued
and determined via $Y=X_{1}\,\text{AND}\,X_{2}$. Then, $p_{Y}(0)=3/4$
and $p_{Y}(1)=1/4$, and both sources have the same channel:
\[
p_{X_{s}\vert Y}(x\vert y)=\begin{cases}
2/3 & \text{if }y=0,x=0\\
1/3 & \text{if }y=0,x=1\\
0 & \text{if }y=1,x=0\\
1 & \text{if }y=1,x=1
\end{cases}
\]
Because the two source channels are the same, the Blackwell redundancy
obeys $\RR=I(Y;X_{1})=I(Y;X_{2})=0.311$ bits \citep{kolchinskyNovelApproachPartial2022}.
From \eqsname(\ref{eq:lb}) and (\ref{eq:up}), we see that $\RBR=\RR$
across all compression rates. In this system, all information provided
by the sources is redundant, so there is no strict tradeoff between
prediction and compression. The RB curve (not shown) consists of a
single point, $(I(Q;Y\vert S),I(Q;S\vert Y))=(0.311,0)$.

\begin{figure*}
\includegraphics[width=1\textwidth]{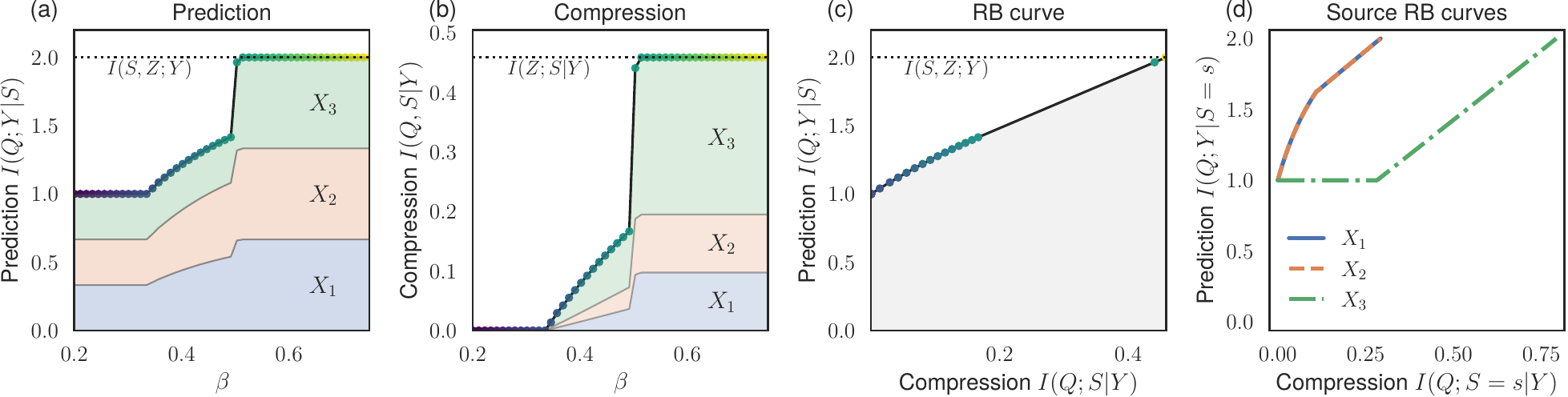}

\caption{\protect\label{fig:bits}RB analysis for the system with
a 3-spin target (Example 4).\textbf{ (a)} and \textbf{(b)} Prediction
and compression values found by optimizing the RB Lagrangian (\ref{eq:opt3})
at different $\beta$. Contributions from individual sources are shown
as shaded regions. \textbf{(c)} The
RB curve shows the tradeoff between optimal compression and prediction
values; marker colors correspond to the $\beta$ values as in (a) and
(b). \textbf{(d)} RB curves for individual sources.}
\end{figure*}

\vspace{5pt}
\noindent \textbf{Example 3.} 
We now consider a more sophisticated example with four sources. The target
is binary-valued and uniformly distributed, $p_{Y}(y)=1/2$ for $y\in\{0,1\}$.
There are four binary-valued sources, where the conditional distribution
of each source $s\in\{1,2,3,4\}$ is a binary symmetric channel with
error probability $\epsilon_{s}$:
\begin{equation}
p_{X_{s}\vert Y}(x\vert y)=\begin{cases}
1-\epsilon_{s} & \text{if }y=x\\
\epsilon_{s} & \text{if }y\ne x
\end{cases}\label{eq:bsc}
\end{equation}
We take $\epsilon_{1}=\epsilon_{2}=0.1$, $\epsilon_{3}=0.2$, and
$\epsilon_{4}=0.5$. Thus, sources $X_{1}$ and $X_{2}$ provide most information about the target; $X_{3}$ provides less information; $X_{4}$  is completely independent of the target.

We performed our RB analysis and plot the RB prediction values in Figure~\ref{fig:4src}(a)
and the compression values in Figure~\ref{fig:4src}(b), as found by
optimizing the RB Lagrangian at different $\beta$. At small $\beta$, the
prediction converges to the Blackwell redundancy, $I(Q;Y\vert S)=\RR=0$,
and there is complete loss of information about source identity, $I(Q;S\vert Y)=0$.
At large $\beta$, the prediction is equal to the maximum $I(Z;Y\vert S)\approx0.335\text{ bit}$,
and compression is equal to $I(Q;S\vert Y)\approx0.104\text{ bit}$.
Figure~\ref{fig:4src}(c) shows the RB curve.

In Figure~\ref{fig:4src}(a) and (b), we show the additive contributions
to the prediction and compression terms from the individual sources, $\sdist(s)I(Q;Y\vert S=s)$
and $\sdist(s)I(Q;S=s\vert Y)$, respectively, as shaded regions. We
also show the resulting RB curves for individual sources in Figure~\ref{fig:4src}(d). 

As expected, source $X_{4}$ does not contribute to the prediction at
any level of compression, in accord with the fact that $I(Q;Y\vert S=s)\le I(X_{4};Y)=0$.
Sources $X_{1}$ and $X_{2}$ provide the same amount of prediction
and compression at all points, up to the maximum $I(X_{1};Y)=I(X_{2};Y)\approx0.531$.
Source $X_{3}$ provides the same amount of prediction and compression
as sources $X_{1}$ and $X_{2}$, until it hits its maximum prediction $I(X_{3};Y)\approx0.278$.
As shown in Figure~\ref{fig:4src}(d), at this point, $X_{3}$ splits
off from sources $X_{1}$ and $X_{2}$ and its compression contribution
decreases to 0; this is compensated by increasing the compression
cost of sources $X_{1}$ and $X_{2}$. The same behavior can also
be seen in Figure~\ref{fig:4src}(a) and (b), where we see that the
solutions undergo phase transitions as different optimal strategies
are uncovered at increasing $\beta$. Importantly, by considering
the prediction/compression contributions from the the individual sources,
we can identify that sources $X_{1}$ and $X_{2}$ provide the most
redundant information.

Let us comment on the somewhat surprising fact that, at
larger $\beta$, the compression cost of $X_{3}$ decreases---even
while its prediction contribution remains constant and the prediction contribution
from $X_{1}$ and $X_{2}$ increases. At first glance, this appears counter-intuitive
if one assumes that, in order to increase prediction from $X_{1}$
and $X_{2}$, the bottleneck channel $p_{Q\vert Y}$ should approach
$p_{X_{1}\vert Y}=p_{X_{2}\vert Y}$, thereby increasing the deficiency
$\delta_{D}(p_{X_{3}\vert Y},p_{Q\vert Y})$ and the compression cost
of $X_{3}$ via the bound (\ref{eq:defbnd}). In fact, this is not
the case, because the prediction is quantified via the conditional mutual
information $I(Q;Y\vert S)$, not the mutual information $I(Q;Y)$. Thus, it is possible
that the prediction contributions from $X_{1}$ and $X_{2}$ are
large, even when the bottleneck channel $p_{Q\vert Y}$ does not closely
resemble $p_{X_{1}\vert Y}=p_{X_{2}\vert Y}$.

More generally, this example shows that it is possible
for the prediction contribution from a given source to stay the same,
or even increase, while its compression cost decreases. In other words,
as can be seen from Figure~\ref{fig:4src}(d), it is possible for
the RB curves of the individual sources to be non-concave and non-monotonic.
It is only the overall RB curve, Figure~\ref{fig:4src}(c), representing the optimal prediction--compression tradeoff on average, that must be concave and monotonic.

\vspace{5pt}
\noindent \textbf{Example 4.} 
In our final example, the target consists of three binary spins with a
uniform distribution, so $Y=(Y_{1},Y_{2},Y_{3})$ and $p_{Y}(y)=1/8$
for all $y$. There are three sources, each of which contains two
binary spins. Sources $X_{1}$ and $X_{2}$ are both equal to the
first two spins of the target $Y$, $X_{1}=X_{2}=(Y_{1},Y_{2})$.
Source $X_{3}$ is equal to the first and last spin of the target,
$X_{3}=(Y_{1},Y_{3})$.

Each source provides $I(Y;X_{s})=2\text{ bits}$ of mutual information
about the target. The Blackwell redundancy $\RR$ is 1 bit, reflecting
the fact that there is a single binary spin that is included in all
sources ($Y_{1}$).

We performed our RB analysis and plot the RB prediction values in Figure~\ref{fig:bits}(a)
and the compression values in Figure~\ref{fig:bits}(b), as found by
optimizing the RB Lagrangian at different $\beta$. At small $\beta$, the
prediction converges to the Blackwell redundancy, $I(Q;Y\vert S)=\RR=1$,
and $I(Q;S\vert Y)=0$. At large $\beta$, the prediction is equal
to the maximum $I(Z;Y\vert S)=2\text{ bit}$, and compression is equal
to $I(Z;S\vert Y)\approx0.459$. Figure~\ref{fig:bits}(c) shows
the RB curve. As in the previous example, the RB curve undergoes phase
transitions as different optimal strategies are uncovered at different
$\beta$. 

In Figure~\ref{fig:bits}(a) and (b), we show the additive contributions
to the prediction and compression terms from the individual sources, $\sdist(s)I(Q;Y\vert S=s)$
and $\sdist(s)I(Q;S=s\vert Y)$, as shaded regions. We also show the
resulting RB curves for individual sources in Figure~\ref{fig:bits}(d).

Observe that sources $X_{1}$ and $X_{2}$ provide more redundant
information at a given level of compression. For instance, as shown in Figure~\ref{fig:bits}(d), at source
compression $I(Q;S=s\vert Y)\approx.25$, $X_{1}$ and $X_{2}$ provide
2 bits of prediction, while $X_{3}$ provides only a single bit. This
again shows how the RB source decomposition can be used for identifying
sources with high levels of redundancy.

\subsection{Continuity}

It is known that the Blackwell redundancy $\RR$ can be discontinuous as a
function of the probability distribution of the target and source
channels \citep{kolchinskyNovelApproachPartial2022}. In Ref.~\citep{kolchinskyNovelApproachPartial2022},
we explain the origin of this discontinuity in geometric terms and
provide sufficient conditions for Blackwell redundancy to be continuous.
Nonetheless, the discontinuity of $\RR$ is sometimes seen as an undesired
property.

On the other hand, as we show in this section, the value of RB is
continuous in the probability distribution for all $R>0$.
\begin{thm}
\label{thm:cont}For finite-dimensional systems and $\rrate>0$, $\RBR$
is a continuous function of the probability values $p_{X_{\vvss}\vert Y}(x\vert y)$,
$p_{Y}(y)$, and $\sdist(s)$.
\end{thm}
Thus, by relaxing the compression constraint in Theorem~\ref{thm:1},
we ``smooth out'' the behavior of Blackwell redundancy and arrive
at a continuous measure. We illustrate this using a simple example.


\vspace{5pt}
\noindent \textbf{Example.} We consider the COPY gate, a standard example in the PID literature.
Here, there are two binary-valued sources jointly distributed according
to 
\[
p_{X_{1}X_{2}}(x_{1},x_{2})=\begin{cases}
1/2-\epsilon/4 & \text{if }x_{1}=x_{2}\\
\epsilon/4 & \text{if }x_{1}\ne x_{2}
\end{cases}
\]
The parameter $\epsilon$ controls the correlation between the two
sources, with perfect correlation at $\epsilon=0$ and complete independence
at $\epsilon=1$. The target $Y$ is a copy of the joint outcome of
the two sources, $Y=(X_{1},X_{2})$.

\begin{figure}
\centering
\includegraphics[width=.7\columnwidth]{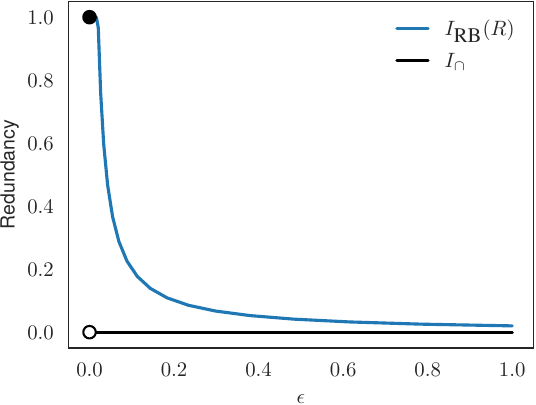}

\caption{\protect\label{fig:cont}The RB function $\protect\RBR$ is continuous
in the underlying probability distribution for $R>0$, while Blackwell
redundancy can be discontinuous. Here illustrated on the COPY gate,
$Y=(X_{1},X_{2})$, as a function of correlation strength $\epsilon$
between $X_{1}$ and $X_{2}$ (perfect correlation at $\epsilon=0$,
independence at $\epsilon=1$). Blackwell redundancy jumps from $\protect\RR=1$
at $\epsilon=0$ to $\protect\RR=0$ at $\epsilon>0$, while $\protect\RBR$
(at $R=0.01$) decays continuously.}
\end{figure}

It is known that Blackwell redundancy $\RR$ is discontinuous for
this system, jumping from $\RR=1$ at $\epsilon=0$ to $\RR=0$ for
$\epsilon>0$ \citep{kolchinskyNovelApproachPartial2022}. On the
other hand, the RB function $\RBR$ is continuous for $R>0$. Figure~\ref{fig:cont}
compares the behavior of Blackwell redundancy and RB as a function
of $\epsilon$, at $R=0.01$ bits. In particular, it can be seen that
$\RBR=1$ at $\epsilon=0$ and then decays continuously as $\epsilon$
increases. 

\section{Iterative Algorithm}

\label{sec:algo}

We provide an iterative algorithm to solve the RB optimization
problem. This algorithm is conceptually similar to the Blahut--Arimoto
algorithm, originally employed for rate distortion problems and later
adapted to solve the original IB problem \citep{tishbyInformationBottleneckMethod1999}.
A Python implementation of our algorithm is available at \url{https://github.com/artemyk/pid-as-ib};
there, we also provide updated code to exactly compute Blackwell redundancy 
(applicable to small systems).

To begin, we consider the RB Lagrangian optimization problem, \eqname(\ref{eq:opt3}).
We rewrite this optimization problem using the KL divergence:
\begin{equation}
F_{\text{RB}}(\cebL)=\max_{r_{Q\vert S\redX}}D(r_{Y\vert QS}\Vert p_{Y\vert S})-\frac{1}{\beta}D(r_{Q\vert SY}\Vert r_{Q\vert Y}).\label{eq:RBlagrangian2}
\end{equation}
Here, notation like $r_{Q\vert Y}$, $r_{Y\vert QS}$, etc., refers
to distributions that include $Q$ and therefore depend on the optimization
variable $r_{Q\vert S\redX}$, while notation like $p_{Y\vert S}$
refers to distributions that do not depend on $Q$ and are not varied
under the optimization. Every choice of conditional distribution $r_{Q\vert S\redX}$
induces a joint distribution $r_{YS\redX Q}=p_{YS\redX}r_{Q\vert S\redX}$
via \eqname(\ref{eq:jointd}).

We can rewrite the first KL term in \eqname(\ref{eq:RBlagrangian2})
as
\begin{align*}
 D(r_{Y\vert QS}\Vert p_{Y\vert S})& =D(r_{Y\vert QS}\Vert p_{Y\vert S})-\min_{\omega_{YS\redX Q}}D(r_{Y\vert QS}\Vert\omega_{Y\vert QS})\\
 & =\max_{\omega_{YS\redX Q}}\mathbb{E}_{p_{YS\redX}r_{Q\vert S\redX}}\left[\ln\frac{\omega(y\vert q,s)}{p(y\vert s)}\right].
\end{align*}
where $\mathbb{E}$ indicates the expectation, and we introduced the variational
distribution $\omega_{YS\redX Q}$. The maximum is achieved by $\omega_{YS\redX Q}=r_{YS\redX Q}$,
which gives $\omega_{Y\vert QS}=r_{Y\vert QS}$. We rewrite the second
KL term in \eqname(\ref{eq:RBlagrangian2}) as
\begin{align*}
D(r_{Q\vert SY}\Vert r_{Q\vert Y}) & =D(r_{Q\vert SY}r_{Z\vert SYQ}\Vert r_{Q\vert Y}r_{Z\vert SYQ})\\
 & =\min_{\omega_{YS\redX Q}}D(r_{Q\vert SY}r_{Z\vert SYQ}\Vert\omega_{Q\vert Y}\omega_{Z\vert SYQ}).
\end{align*}
Here, we introduce the variational distribution $\omega_{YS\redX Q}$,
where the minimum is achieved by $\omega_{YS\redX Q}=r_{YS\redX Q}$.
The term $r_{Q\vert SY}r_{Z\vert SYQ}$ can be rewritten as 
\begin{multline*}
r(q\vert s,y)r(z\vert s,y,q)  =\frac{r(z,s,y,q)}{p(s,y)}  \\=\frac{r(q\vert s,z)p(z,s,y)}{p(s,y)}=r(q\vert s,z)p(z\vert s,y)
\end{multline*}
where we used the Markov condition $Q-(S,\redX)-Y$. In this way,
we separate the contribution from the conditional distribution $r_{Q\vert S\redX}$
being optimized. 

Combining the above allows us to rewrite \eqname(\ref{eq:RBlagrangian2})
as
\begin{multline}
F_{\text{RB}}(\cebL)=\max_{r_{Q\vert\redX S},\omega_{YS\redX Q}}\mathbb{E}_{p_{YS\redX}r_{Q\vert S\redX}}\left[\ln\frac{\omega(y\vert q,s)}{p(y\vert s)}\right]\\-\frac{1}{\cebL}D(r_{Q\vert S\redX}p_{Z\vert SY}\Vert\omega_{Q\vert Y}\omega_{Z\vert SYQ}).\label{eq:iterative-obj}
\end{multline}
We now optimize this objective in an iterative and alternating manner
with respect to $r_{Q\vert S\redX}$ and $\omega_{YS\redX Q}$. Formally,
let $\mathcal{L}(r_{Q\vert S\redX},\omega_{YS\redX Q})$ refer to
the objective in \eqname(\ref{eq:iterative-obj}). Then, starting from
some initial guess $r_{Q\vert S\redX}^{(0)}$, we generate a sequence
of solutions
\begin{align}
\omega_{YS\redX Q}^{(t+1)} & =\argmax_{\omega_{YS\redX Q}}\mathcal{L}(r_{Q\vert S\redX}^{(t)},\omega_{YS\redX Q})\label{eq:optA-1}\\
r_{Q\vert S\redX}^{(t+1)} & =\argmax_{r_{Q\vert S\redX}}\mathcal{L}(r_{Q\vert S\redX},\omega_{YS\redX Q}^{(t+1)})\label{eq:optB-1}
\end{align}
Each optimization problem can be solved in closed form. As already
mentioned, the optimizer in \eqname(\ref{eq:optA-1}) is
\[
\omega_{YS\redX Q}^{(t+1)}=r_{YS\redX Q}^{(t)}=r_{Q\vert S\redX}^{(t)}p_{SZY}.
\]
The optimization (\ref{eq:optB-1}) can be solved by taking derivatives,
giving
\begin{align*}
r^{(t+1)}(q\vert s,z)\propto e^{\sum_{y}p(y\vert s,z)\Big[\cebL\ln\omega^{(t)}(y\vert q,s)-\ln\frac{p(z\vert s,y)}{\omega^{(t)}(q\vert y)\omega^{(t)}(z\vert s,y,q)}\Big]}
\end{align*}
where the proportionality constant in $\propto$ is fixed by normalization
$\sum_{q}r^{(t+1)}(q\vert s,z)=1$.

Each iteration increases the value of objective $\mathcal{L}$. Since
the objective is upper bounded by $I(\redX;Y\vert S)$, the algorithm
is guaranteed to converge. However, as in the case of the original
IB problem, the objective is not jointly convex in both arguments,
so the algorithm may converge to a local maximum or a saddle point,
rather than a global maximum. This can be partially alleviated by
running the algorithm several times starting from different initial
guesses $r_{Q\vert SZ}^{(0)}$.

When the RB is not strictly concave, it is more appropriate to optimize
the exponential RB Lagrangian (\ref{eq:RBlagrangianExp}) or another
objective that combines the prediction and compression terms in a nonlinear
manner \citep{kolchinskyCaveatsInformationBottleneck2019,rodriguez2020convex}.
The algorithm described above can be used with such objectives after
a slight modification. For instance, for the exponential RB Lagrangian,
we modify (\ref{eq:iterative-obj}) as
\begin{multline}
F_{\text{RB}}^{\exp}(\cebL)=\max_{r_{Q\vert S\redX},\omega_{YS\redX Q}}\mathbb{E}_{p_{YS\redX}r_{Q\vert S\redX}}\left[\ln\frac{\omega(y\vert q,s)}{p(y\vert s)}\right]\\-\frac{1}{\beta}e^{D(r_{Q\vert ZS}p_{Z\vert SY}\Vert\omega_{Q\vert Y}\omega_{Z\vert SYQ})}.
\end{multline}
A similar analysis as above leads to the following iterative optimization
scheme:
\begin{align*}
  &\omega_{YS\redX Q}^{(t+1)}=r_{Q\vert S\redX}^{(t)}p_{SZY}\\
  &r^{(t+1)}(q\vert s,z)\propto\\
  & \qquad   e^{\sum_{y}p(y\vert s,z)\Big[\beta^{(t)}\ln\omega^{(t)}(y\vert q,s)-\ln\frac{p(z\vert s,y)}{\omega^{(t)}(q\vert y)\omega^{(t)}(z\vert s,y,q)}\Big]},
\end{align*}
where $\cebL^{(t)}=\cebL e^{-I_{r^{(t)}}(Q;S\vert Y)}$ is an effective
inverse temperature. (Observe that, unlike the squared Lagrangian
\citep{kolchinskyCaveatsInformationBottleneck2019}, the exponential
Lagrangian leads to an effective inverse temperature $\cebL^{(t)}$ that is always finite and converges
to $\beta$ as $I_{r^{(t)}}(Q;S\vert Y)\to0$.)

When computing an entire RB curve, as in Figure~\ref{fig:unq}(a)-(c),
we found good results by annealing, that is by re-using the optimal
$r_{Q\vert S\redX}$ found for one $\beta$ as the initial guess at
higher $\beta$. For quantifying the value of the RB
function $\RBR$ at a fixed $R$, as in Figure~\ref{fig:cont}, we
approximated $\RBR$ via a linear interpolation of the RB prediction
and compression values recovered from the RB Lagrangian at varying
$\beta$.

\section{Discussion}

\label{sec:Discussion}

In this paper, we propose a generalization of Blackwell redundancy,
termed the redundancy bottleneck (RB), formulated as an information-bottleneck-type
tradeoff between prediction and compression. We studied some implications
of this formulation and proposed an efficient numerical algorithm to
solve the RB optimization problem.

We briefly mention some directions for future work.

The first direction concerns our iterative algorithm. The algorithm
is only applicable to systems where it is possible to enumerate the
outcomes of the joint distribution $p_{QYSZ}$. This is impractical
for discrete-valued variables with very many outcomes, as well as continuous-valued
variables as commonly found in statistical and machine learning
settings. In future work, it would be useful to develop RB algorithms
suitable for such datasets, possibly by exploiting the kinds of variational
techniques that have recently gained traction in machine learning
applications of IB \citep{alemi2016deep,kolchinsky2019nonlinear,fischer2020conditional}.

The second direction would explore connections between the RB and
other information-theoretic objectives for representation learning.
To our knowledge, the RB problem is novel to the literature. However,
it has some similarities to existing objectives, including among others the conditional entropy
bottleneck \citep{fischer2020conditional}, multi-view IB \citep{federici2020learning},
and the privacy funnel and its variants \citep{makhdoumi2014information}. Showing formal connections between these objectives
would be of theoretical and practical interest, and could lead to
new interpretations of the concept of PID redundancy.

Another direction would explore the relationship between RB and information-theoretic
measures of causality \citep{janzing2013quantifying,ay2021confounding}.
In particular, if the different sources represent some
mutually exclusive conditions---such as the age group example provided
at the end of Section \ref{sec:background}---then redundancy could
serve as a measure of causal information flow that is invariant to
the conditioning variable.

Finally, one of the central ideas of this paper is to treat the identity
of the source as a random variable in its own right, which allows us to
consider what information different bottleneck variables reveal about
the source. In this way, we convert the search for topological or 
combinatorial structure in multivariate systems into an interpretable
and differential information-theoretic objective. This technique may
be useful in other problems that consider how information is distributed
among variables in complex systems, including other PID measures such as synergy \citep{kolchinskyNovelApproachPartial2022},
information-theoretic measures of modularity \citep{kolchinsky2011prediction,hidaka2018fast},
and measures of higher-order dependency \citep{rosas2016understanding,rosas2019quantifying}.

\vspace{6pt}

\begin{acknowledgments}
I thank Nihat Ay, Daniel Polani, Fernando Rosas, and especially, André
Gomes for useful feedback. I also thank the organizers of the ``Decomposing
Multivariate Information in Complex Systems'' (DeMICS 23) workshop
at the Max Planck Institute for the Physics of Complex Systems (Dresden,
Germany), which inspired this work. This project has received funding
from the European Union’s Horizon 2020 research and innovation programme
under the Marie Skłodowska-Curie Grant Agreement No. 101068029.
\end{acknowledgments}

\section*{Appendix}
\label{app:1}
We provide proofs of the theorems in the main text. Throughout, we
use $D$ for the Kullback--Leibler (KL) divergence, $H$ for the Shannon entropy,
and $I$ for the mutual information.

\subsection*{Proof of Theorem~\ref{thm:1}}

We begin by proving a slightly generalized version of Theorem~\ref{thm:1}, that is we show the equivalence between the two optimization problems:
\begin{align}
I_{\cap}&=\sup_{Q}\,I(Q;Y)\where Q\preceq_{Y}X_{\vvss}\;\forall\vvss\label{eq:opt1-app}\\
I_{\cap}&=\sup_{Q:Q-(\redX,S)-Y}I(Q;Y\vert S)\where I(Q;S\vert Y)=0.\label{eq:thm1-app}
\end{align}
The slight generalization comes from replacing $\max$ by $\sup$, so that the result also holds for systems with infinite-dimensional sources, where the supremum is not guaranteed to be achieved. For finite-dimensional systems, the supremum is always achieved, and we reduce to the simpler case of \eqsname\eqref{eq:opt1} and \eqref{eq:thm1}.

\begin{proof}
Let $V_{1}$ indicate the supremum in \eqname(\ref{eq:opt1-app}) and $V_{2}$
the supremum in \eqname(\ref{eq:thm1-app}), given some
$\sdist(s)$ with full support. We prove that $V_{1}=V_{2}$.

We will use that, for any distribution that has the form
of \eqname(\ref{eq:jointd}) and obeys $I(Q;S\vert Y)=0$, the following
holds:
\begin{equation}
\begin{aligned}I(Q;Y\vert S) & =H(Q\vert S)-H(Q\vert S,Y)\\
 & =H(Q)-H(Q\vert Y)=I(Q;Y)
\end{aligned}
\label{eq:eq0}
\end{equation}
Here, we used the Markov condition $Q-Y-S$, as well as $I(Q;S)=H(Q)-H(Q\vert S)=0$,
as follows from \eqname(\ref{eq:YSind}) and the data-processing inequality.

Let $Q$ be a feasible random variable that comes within $\epsilon\ge0$
of the objective in (\ref{eq:opt1-app}), $I(Q;Y)\ge V_{1}-\epsilon$.
Define the joint distribution:
\[
p_{QYS\redX}(q,y,s,\redx)=\kappa_{Q\vert X_{s}}(q\vert\redx)p_{Y}(y)\sdist(\vvss)p_{X_{s}\vert Y}(\redx\vert y)
\]
whenever $\redx\in\mathcal{X}_{s}$, otherwise $p_{QYS\redX}(q,y,s,\redx)=0$.
Here, we used the channels $\kappa_{Q\vert X_{s}}$ associated with
the Blackwell relation $Q\preceq_{Y}X_{\vvss}$, so that $p_{Q\vert Y}=\kappa_{Q\vert X_{s}}\circ p_{X_{s}\vert Y}$.
Under the distribution $p_{QYS\redX}$, the Markov conditions $Q-(S,\redX)-Y$
and $Q-Y-S$ hold, the latter since 
\begin{equation}
p_{QS\vert Y}(q,s\vert y)=p_{Q\vert Y}(q\vert y)\sdist(s).\label{eq:nnn}
\end{equation}
Therefore, this distribution has the form of \eqsname(\ref{eq:jointd})
and (\ref{eq:pdef}) and satisfies the constraints in \eqname(\ref{eq:thm1-app}).
Using \eqname(\ref{eq:eq0}), we then have 
\[
V_{1}-\epsilon\le I(Q;Y)=I(Q;Y\vert S)\le V_{2}.
\]

Conversely, let $p_{YS\redX Q}$ be a feasible joint distribution
for the optimization of \eqname(\ref{eq:thm1-app}) that comes within $\epsilon\ge0$
of the supremum, $I(Q;Y\vert S)\ge V_{2}-\epsilon$. Using the form
of this joint distribution from \eqname(\ref{eq:jointd}), we can write
\begin{align*}
p_{Q\vert Y}(q\vert y) & \stackrel{(a)}{=}p_{Q\vert YS}(q\vert y,s)\\
 & =\sum_{\redx}p_{Q\vert YS\redX}(q\vert y,s,\redx)p_{\redX\vert YS}(\redx\vert y,s)\\
 & \stackrel{(b)}{=}\sum_{\redx}p_{Q\vert S\redX}(q\vert\redx,\vvss)p_{\redX\vert YS}(\redx\vert y,s)\\
 & \stackrel{(c)}{=}\sum_{\redx}p_{Q\vert S\redX}(q\vert\redx,\vvss)p_{X_{s}\vert Y}(\redx\vert y)
\end{align*}
In $(a)$, we used $I(Q;S\vert Y)=0$; in $(b)$, we used $Q-(S,\redX)-Y$;
in $(c)$, we used that $p_{\redX\vert YS=s}=p_{X_{s}\vert Y}$.
This implies that $p_{Q\vert Y}\preceq p_{X_{s}\vert Y}$ for all
$s$. Therefore, $p_{Q\vert Y}$ satisfies the constraints in \eqname(\ref{eq:opt1-app}),
so $I(Q;Y)\le V_{1}$. Combining with \eqname(\ref{eq:eq0}) implies
\[
V_{2}-\epsilon\le I(Q;Y\vert S)=I(Q;Y)\le V_{1}.
\]

Taking the limit $\epsilon\to0$ gives the desired result.
\end{proof}

\subsection*{Proof of Theorem~\ref{thm:conc}}

We now prove a slightly generalized version of Theorem~\ref{thm:conc}. We show that the solution to the following optimization problem is non-decreasing and concave in $\rrate$:
\begin{equation}
\RBR:=\quad\sup_{\mathclap{Q:Q-(\redX,S)-Y}}\quad I(Q;Y\vert S)\where I(Q;S\vert Y)\le\rrate.\label{eq:opt2-sup}
\end{equation}
The slight generalization comes from replacing $\max$ in \eqname\eqref{eq:opt2-1} by $\sup$, so that the result also holds for systems with infinite-dimensional sources where the supremum is not guaranteed to be  achieved.

\begin{proof} 
$\RBR$ is non-decreasing in $\rrate$ because larger $\rrate$ give
weaker constraints (larger feasible set) in the maximization problem
\eqref{eq:opt2-sup}. 

To show concavity, consider any two
points on the RB curve as defined by \eqname\eqref{eq:opt2-sup}: $(\rrate,\RBR)$ and $(\rrate^{\prime},\RB{\rrate^{\prime}})$.
For any $\epsilon >0$, there exist $Q$ and $Q^\prime$ such that 
\begin{align*}
I(Q;S\vert Y)&\le\rrate & I(Q;Y\vert S) &\ge \RBR-\epsilon \\
I(Q^\prime;S\vert Y)&\le\rrate^{\prime}  &I(Q^\prime;Y\vert S) &\ge \RB{\rrate^{\prime}}-\epsilon
\end{align*}
Without
loss of generality, suppose that both variables have the same
set of outcomes $\mathcal{Q}$. Then, we define a new random variable
$Q_{\lambda}$ with outcomes $\mathcal{Q}_{\lambda}=\{1,2\}\times\mathcal{Q}$,
as well as a family of conditional distributions parameterized by
$\lambda\in[0,1]$:
\begin{align*}
p_{Q_{\lambda}\vert\redX S}(1,q\vert\redx,s) & =\lambda p_{Q\vert\redX S}(q\vert\redx,s)\\
p_{Q_{\lambda}\vert\redX S}(2,q\vert\redx,s) & =(1-\lambda)p_{Q^{\prime}\vert\redX S}(q\vert\redx,s)
\end{align*}
In this way, we define $Q_{\lambda}$ via a disjoint convex mixture
of $Q$ and $Q^{\prime}$ onto non-overlapping subspaces, with $\lambda$
being the mixing parameter. With a bit of algebra, it can be verified
that, for every $\lambda$,
\[
H(Q_{\lambda}\vert Y)=\lambda H(Q\vert Y)+(1-\lambda)H(Q^{\prime}\vert Y),
\]
and similarly for $H(Q_{\lambda}\vert Y,S)$ and $H(Q_{\lambda}\vert S)$.
Therefore,
\begin{align*}
I(Q_{\lambda};S\vert Y) & =\lambda I(Q;S\vert Y)+(1-\lambda)I(Q^{\prime};S\vert Y)\\
 & \le \lambda\rrate+(1-\lambda)\rrate^{\prime}\\
I(Q_{\lambda};Y\vert S) & =\lambda I(Q;Y\vert S)+(1-\lambda)I(Q^{\prime};Y\vert S)\\
 & \ge\lambda\RBR+(1-\lambda)\RB{\rrate^{\prime}} -\epsilon
\end{align*}
Since $\RBbase$ is defined via a maximization, we have 
\begin{multline*}
\RB{\lambda\rrate+(1-\lambda)\rrate^{\prime}}  \ge I(Q_{\lambda};Y\vert S) \\\ge \lambda\RBR+(1-\lambda)\RB{\rrate^{\prime}}-\epsilon.
\end{multline*}
Taking the limit $\epsilon \to 0$ proves the concavity.
\end{proof}

\subsection*{Proof of Theorem~\ref{thm:card}}
\begin{proof}
We show that, for any $Q$ that achieves $I(Q;S\vert Y)\le\rrate$,
there is another $Q^{\prime}$ with cardinality $\left|\mathcal{Q}^{\prime}\right|\le\sum_{\vvss}\left|\mathcal{X}_{\vvss}\right|+1$
that satisfies $I(Q^{\prime};S\vert Y)\le\rrate$ and $I(Q^{\prime};Y\vert S)\ge I(Q;Y\vert S)$.

Consider any joint distribution $p_{QS\redX Y}$ from \eqname(\ref{eq:jointd})
that achieves $I(Q;S\vert Y)\le\rrate$, and let $\mathcal{Q}$ be the corresponding set of outcomes of $Q$. Fix the corresponding conditional
distribution $p_{S\redX\vert Q}$, and note that it also determines
the conditional distributions:
\begin{align}
p_{YS\redX\vert Q}(y,s,\redx\vert q) & =p_{Y\vert S\redX}(y\vert s,\redx)p_{S\redX\vert Q}(s,\redx\vert q)\\
 & =\frac{\sdist(s)p_{X_{s}\vert Y}(\redx\vert y)p_{Y}(y)}{p_{S\redX}(s,\redx)}p_{S\redX\vert Q}(s,\redx\vert q)\label{eq:appmmm1}\\
p_{Y\vert SQ}(y\vert s,q) & =\frac{\sum_{\redx}p_{YS\redX\vert Q}(y,s,\redx\vert q)}{\sum_{\redx,y^{\prime}}p_{YS\redX\vert Q}(y^{\prime},s,\redx\vert q)}\\
p_{S\vert YQ}(s\vert y,q) & =\frac{\sum_{\redx}p_{YS\redX\vert Q}(y,s,\redx\vert q)}{\sum_{\redx,s^{\prime}}p_{YS\redX\vert Q}(y,s^{\prime},\redx\vert q)}
\end{align}

Next, consider the following linear program:
\begin{align}
V =\max_{\omega_{Q^{\prime}}\in\Delta}& \sum_{q}\omega_{Q^{\prime}}(q)D(P_{Y\vert SQ=q}\Vert P_{Y\vert S})\label{eq:v}\\
\text{where\,} & \sum_{q}\omega_{Q^{\prime}}(q)p_{S\redX\vert Q}(s,\redx\vert q)=p_{S\redX}(s,\redx)\;\forall s,\redx\label{eq:fs}\\
 & \sum_{q}\omega_{Q^{\prime}}(q)H({S\vert Y,Q=q})=H({S\vert Y,Q})\label{eq:lastc}
\end{align}
where $\Delta$ is the $\vert\mathcal{Q}\vert$-dimensional unit simplex,
and we use the notation $H({S\vert Y,Q=q})=-\sum_{y,s}p_{YS\vert Q}(y,s\vert q)\ln p_{S\vert YQ}(s\vert y,q)$.
The first set of constraints (\ref{eq:fs}) guarantees that $\omega_{Q^{\prime}}p_{YS\redX\vert Q}$
belongs to the family (\ref{eq:jointd}) and, in particular, that the marginal
over $(S,\redX,Y)$ is $\sdist(s)p_{X_{s}\vert Y}(\redx\vert y)p_{Y}(y)$ (see \eqname(\ref{eq:appmmm1})).
There are $\sum_{\vvss}\left|\mathcal{X}_{\vvss}\right|$ possible outcomes
of $(s,\redx)$, but $\sum_{s,\redx}p_{S\redX}(s,\redx)=1$ by the conservation
of probability. Therefore, \eqname(\ref{eq:fs}) effectively imposes $\sum_{\vvss}\left|\mathcal{X}_{\vvss}\right|-1$
constraints. The last constraint (\ref{eq:lastc}) guarantees that
$H({S\vert Y,Q^{\prime}})=H({S\vert Y,Q})$; hence, 
\begin{align*}
I(Q^{\prime};S\vert Y) & =H({S\vert Y})-H({S\vert Y,Q^{\prime}})\\
 & =H({S\vert Y})-H({S\vert Y,Q})=I(Q;S\vert Y)\le\rrate.
\end{align*}

\eqname(\ref{eq:v}) involves a maximization of a linear function over
the simplex, subject to $\sum_{\vvss}\left|\mathcal{X}_{\vvss}\right|$
hyperplane constraints. The feasible set is compact, and the maximum
is achieved at one of the extreme points of the feasible set.
By Dubin’s theorem~\citep{dubins1962extreme}, any extreme point
of this feasible set can be expressed as a convex combination of at
most $\sum_{\vvss}\left|\mathcal{X}_{\vvss}\right|+1$ extreme points
of $\Delta$. Thus, the maximum is achieved by a marginal distribution
$\omega_{Q^{\prime}}$ with support on at most $\sum_{\vvss}\left|\mathcal{X}_{\vvss}\right|+1$
outcomes. This distribution satisfies:
\begin{align*}
\sum_{q}\omega_{Q^{\prime}}(q)D(P_{Y\vert SQ=q}\Vert P_{Y\vert S})\ge\sum_{q}p_{Q}(q)D(P_{Y\vert SQ=q}\Vert P_{Y\vert S})
\end{align*}
since the actual marginal distribution $p_{Q}$ is an element of the feasible set.
Finally, note that 
\begin{align*}
\sum_{q}\omega_{Q^{\prime}}(q)D(P_{Y\vert SQ=q}\Vert P_{Y\vert S}) & =I(Q^{\prime};Y\vert S)\\
\sum_{q}p_{Q}(q)D(P_{Y\vert SQ=q}\Vert P_{Y\vert S}) & =I(Q;Y\vert S);
\end{align*}
therefore, $I(Q^{\prime};Y\vert S)\ge I(Q;Y\vert S)$.
\end{proof}

\subsection*{Proof of Theorem~\ref{thm:cont}}

\global\long\def\odd{v}%
\global\long\def\mand{\omega}%
\global\long\def\lparam{\lambda}%
\global\long\def\mandl{\mand^{\lparam}}%

\begin{proof}
For a finite-dimensional system, we may restrict the optimization
problem in Theorem~\ref{thm:1} to $Q$ with cardinality $\vert\mathcal{Q}\vert\le\sum_{\vvss}\left|\mathcal{X}_{\vvss}\right|+1$
(Theorem~\ref{thm:card}). In this case, the feasible set can be
restricted to a compact set, and the objective is continuous; therefore,
the maximum will be achieved.

\global\long\def\ix{k}%
Now, consider a tuple of random variables $(S,\redX,Y,Q)$ that obey
the Markov conditions $S-Y-\redX$ and $Q-(S,\redX)-Y$. Suppose that
$Q$ achieves the maximum in Theorem~\ref{thm:1} for a given $R>0$:
\begin{equation}
I(Q;Y\vert S)=\RBR,\qquad I(Q;S\vert Y)\le\rrate.\label{eq:defs0}
\end{equation}
Consider also a sequence of random variables $(S_{\ix},\redX_{\ix},Y_{\ix},Q_{k})$
for $k=1,2,3\dots$ where each tuple has the same outcomes
as $(S,\redX,Y,Q)$ and obeys the Markov conditions $S_{k}-Y_{k}-\redX_{k}$
and $Q_{k}-(S_{k},\redX_{k})-Y_{k}$. Let $I_{\mathrm{\text{RB}}}^{\ix}(R)$
indicate the redundancy bottleneck defined in Theorem~\ref{thm:1}
for random variables $\redX_{k},Y_{k},S_{k}$, and suppose that $Q_{k}$
achieves the optimum for problem $\ix$:
\begin{equation}
I(Q_{\ix};Y_{\ix}\vert S_{\ix})=I_{\mathrm{\text{RB}}}^{\ix}(R)\qquad I(Q_{\ix};S_{\ix}\vert Y_{\ix})\le\rrate.\label{eq:defs}
\end{equation}
To prove continuity, we assume that the joint distribution of $(S_{\ix},\redX_{\ix},Y_{\ix})$
approaches the joint distribution of $(S,\redX,Y)$,
\[
\lim_{k}\left\Vert p_{S_{k}\redX_{k}Y_{k}}-p_{SXY}\right\Vert _{1}=0.
\]

We first show that 
\begin{equation}
\RBR\ge\lim_{k}I_{\mathrm{\text{RB}}}^{\ix}(R).\label{eq:lb0}
\end{equation}
First, observe that given our assumption that $p_{S\redX}$ has full support, we can always
take $k$ sufficiently large so that each $p_{S_{\ix}\redX_{\ix}}$
has full support. Next, we define the random variable $Q_{k}^{\prime}$ that obeys the Markov
condition $Q_{\ix}^{\prime}-(S,\redX)-Y$, with conditional distribution:
\[
p_{Q_{\ix}^{\prime}\vert S\redX}(q\vert s,\redx):=p_{Q_{\ix}\vert S_{\ix}\redX_{\ix}}(q\vert s,\redx).
\]
This conditional distribution is always well-defined, given that $p_{S_{\ix}\redX_{\ix}}$
has the same support as $p_{S\redX}$. By assumption, $p_{S_{\ix}\redX_{\ix}Y_{\ix}}\to p_{S\redX Y}$;
therefore, 
\begin{equation}
p_{Q_{\ix}^{\prime}S\redX Y}-p_{Q_{\ix}S_{\ix}\redX_{\ix}Y_{\ix}}\to0.\label{eq:limlim}
\end{equation}
Conditional mutual information is (uniformly) continuous due to the (uniform)
continuity of entropy \citep[Theorem 17.3.3, ][]{cover_elements_2006}.
Therefore,
\begin{align}
0 =\lim_{k}[I(Q_{\ix};S_{\ix}\vert Y_{\ix})-I(Q_{\ix}^{\prime};S\vert Y)] \le R-\lim_{k}I(Q_{\ix}^{\prime};S\vert Y),\label{eq:lim33}
\end{align}
where we used \eqname(\ref{eq:defs}). We also define another random
variable $Q_{k}^{\alpha}$, which also obeys the Markov condition $Q_{\ix}^{\alpha}-(S,\redX)-Y$,
whose conditional distribution is defined in terms of the convex mixture:
\begin{align}
p_{Q_{k}^{\alpha}\vert S\redX}(q\vert s,\redx) & :=\alpha_{\ix}p_{Q^{\prime}\vert SX}(q\vert s,\redx)+(1-\alpha_{\ix})p_{U}(q),\nonumber \\
\alpha_{\ix} & :=\min\left\{ 1,\frac{R}{I(Q_{\ix}^{\prime};S\vert Y)}\right\} \in[0,1]\label{eq:alphadef}
\end{align}
Here, $p_{U}(q)=1/\vert\mathcal{Q}\vert$ is a uniform distribution
over an auxiliary independent random variable $U$ with outcomes $\mathcal{Q}$.
From the convexity of conditional mutual information \citep{timo2012lossy},
\begin{align*}
I(Q_{\ix}^{\alpha};S\vert Y) & \le\alpha_{\ix}I(Q_{\ix}^{\prime};S\vert Y)+(1-\alpha_{\ix})I(U;S\vert Y)\le R.
\end{align*}
In the last inequality, we used $I(U;S\vert Y)=0$ and plugged in
the definition of $\alpha_{\ix}$. Observe that the random variable
$Q_{k}^{\alpha}$ falls in the feasible set of the maximization problem
that defines $\RBbase$, so
\begin{equation}
\RBR\ge I(Q_{k}^{\alpha};Y\vert S).\label{eq:n1}
\end{equation}
Combining \eqsname(\ref{eq:lim33}), (\ref{eq:alphadef}), and $R>0$
implies that $\alpha_{\ix}\to1$, so 
$$p_{Q_{\ix}^{\alpha}S\redX Y} - p_{Q_{\ix}^{\prime}S\redX Y} \to 0.$$
Combining with \eqsname(\ref{eq:n1}), (\ref{eq:limlim}), and (\ref{eq:defs}),
along with continuity of conditional mutual information, gives
\begin{multline*}
\RBR  \ge\lim_{k}I(Q_{k}^{\alpha};Y\vert S) =\lim_{k}I(Q_{k}^{\prime};Y\vert S) \\
=\lim_{k}I(Q_{k};Y_{k}\vert S_{\ix})=\lim_{k}I_{\mathrm{\text{RB}}}^{\ix}(R).
\end{multline*}

We now proceed in a similar way to prove
\begin{equation}
\RBR\le\lim_{k}I_{\mathrm{\text{RB}}}^{\ix}(R).\label{eq:lb0-1}
\end{equation}
We define the random variable $Q^{\prime\prime}$ that obeys the Markov
condition $Q^{\prime\prime}-(S_{\ix},\redX_{\ix})-Y_{\ix}$, with
conditional distribution
\[
p_{Q_{\ix}^{\prime\prime}\vert S_{\ix}\redX_{\ix}}(q\vert s,\redx)=p_{Q\vert SX}(q\vert s,\redx).
\]
Since $p_{S_{\ix}\redX_{\ix}Y_{\ix}}\to p_{S\redX Y}$ by assumption,
\begin{equation}
p_{Q_{\ix}^{\prime\prime}S_{\ix}\redX_{\ix}Y_{\ix}}-p_{QS\redX Y}\to0.\label{eq:limlim2}
\end{equation}
We then have
\begin{align}
0 & =\lim_{k}[I(Q;S\vert Y)-I(Q_{\ix}^{\prime\prime};S_{\ix}\vert Y_{\ix})] \le R-\lim_{k}I(Q_{\ix}^{\prime\prime};S_{\ix}\vert Y_{\ix}).\label{eq:lim44}
\end{align}
where we used \eqname(\ref{eq:defs0}). We also define the random variable
$Q^{\alpha\prime}$ that obeys the Markov condition $Q^{\alpha\prime}-(S_{\ix},\redX_{\ix})-Y_{\ix}$,
with conditional distribution
\begin{align}
p_{Q_{k}^{\alpha\prime}\vert S_{\ix}\redX_{\ix}}(q\vert s,\redx) & =\alpha_{\ix}^{\prime}p_{Q^{\prime\prime}\vert S_{\ix}\redX_{\ix}}(q\vert s,\redx)+(1-\alpha_{\ix}^{\prime})p_{U}(q),\nonumber \\
\alpha_{\ix}^{\prime} & :=\min\left\{ 1,\frac{R}{I(Q_{\ix}^{\prime\prime};S_{\ix}\vert Y_{\ix})}\right\} \in[0,1]\label{eq:alphadef2}
\end{align}
Using the convexity of conditional mutual information, $I(U;S_{\ix};Y_{\ix})=0$,
and the definition of $\alpha_{\ix}^{\prime}$, we have
\[
I(Q_{\ix}^{\alpha\prime};S_{\ix}\vert Y_{\ix})\le\alpha_{\ix}^{\prime}I(Q_{\ix}^{\prime\prime};S_{\ix}\vert Y_{\ix})+(1-\alpha_{\ix}^{\prime})I(U;S_{\ix}\vert Y_{\ix})\le R.
\]
Therefore, the random variable $Q_{k}^{\alpha\prime}$ falls in the
feasible set of the maximization problem that defines $I_{\mathrm{\text{RB}}}^{\ix}$,
so
\begin{equation}
I_{\mathrm{\text{RB}}}^{\ix}(\rrate)\ge I(Q_{k}^{\alpha\prime};Y_{\ix}\vert S_{k}).\label{eq:up0}
\end{equation}
Combining \eqsname(\ref{eq:lim44}), (\ref{eq:alphadef2}), and $R>0$
implies $\alpha_{\ix}^{\prime}\to1$; therefore, 
$$p_{Q_{\ix}^{\alpha\prime}S_{\ix}\redX_{\ix}Y_{\ix}}-p_{Q_{\ix}^{\prime\prime}S_{\ix}\redX_{\ix}Y_{\ix}}\to0.$$
Combining this with \eqsname(\ref{eq:up0}), (\ref{eq:limlim2}), and (\ref{eq:defs0}),
along with the continuity of conditional mutual information, gives
\begin{multline*}
\lim_{\ix}I_{\mathrm{\text{RB}}}^{\ix}(\rrate)\ge \lim_{\ix}I(Q_{k}^{\alpha\prime};Y_{\ix}\vert S_{k})
\\= \lim_{k}I(Q_{k}^{\prime\prime};Y_{k}\vert S_{\ix}) = I(Q;Y\vert S)=\RBR.
\end{multline*}
\end{proof}

\bibliography{writeup}

\begin{thebibliography}{58}%
\makeatletter
\providecommand \@ifxundefined [1]{%
 \@ifx{#1\undefined}
}%
\providecommand \@ifnum [1]{%
 \ifnum #1\expandafter \@firstoftwo
 \else \expandafter \@secondoftwo
 \fi
}%
\providecommand \@ifx [1]{%
 \ifx #1\expandafter \@firstoftwo
 \else \expandafter \@secondoftwo
 \fi
}%
\providecommand \natexlab [1]{#1}%
\providecommand \enquote  [1]{``#1''}%
\providecommand \bibnamefont  [1]{#1}%
\providecommand \bibfnamefont [1]{#1}%
\providecommand \citenamefont [1]{#1}%
\providecommand \href@noop [0]{\@secondoftwo}%
\providecommand \href [0]{\begingroup \@sanitize@url \@href}%
\providecommand \@href[1]{\@@startlink{#1}\@@href}%
\providecommand \@@href[1]{\endgroup#1\@@endlink}%
\providecommand \@sanitize@url [0]{\catcode `\\12\catcode `\$12\catcode
  `\&12\catcode `\#12\catcode `\^12\catcode `\_12\catcode `\%12\relax}%
\providecommand \@@startlink[1]{}%
\providecommand \@@endlink[0]{}%
\providecommand \url  [0]{\begingroup\@sanitize@url \@url }%
\providecommand \@url [1]{\endgroup\@href {#1}{\urlprefix }}%
\providecommand \urlprefix  [0]{URL }%
\providecommand \Eprint [0]{\href }%
\providecommand \doibase [0]{https://doi.org/}%
\providecommand \selectlanguage [0]{\@gobble}%
\providecommand \bibinfo  [0]{\@secondoftwo}%
\providecommand \bibfield  [0]{\@secondoftwo}%
\providecommand \translation [1]{[#1]}%
\providecommand \BibitemOpen [0]{}%
\providecommand \bibitemStop [0]{}%
\providecommand \bibitemNoStop [0]{.\EOS\space}%
\providecommand \EOS [0]{\spacefactor3000\relax}%
\providecommand \BibitemShut  [1]{\csname bibitem#1\endcsname}%
\let\auto@bib@innerbib\@empty
\bibitem [{\citenamefont {Williams}\ and\ \citenamefont
  {Beer}(2010)}]{williams2010nonnegative}%
  \BibitemOpen
  \bibfield  {author} {\bibinfo {author} {\bibfnamefont {P.~L.}\ \bibnamefont
  {Williams}}\ and\ \bibinfo {author} {\bibfnamefont {R.~D.}\ \bibnamefont
  {Beer}},\ }\bibfield  {title} {\bibinfo {title} {Nonnegative decomposition of
  multivariate information},\ }\href@noop {} {\bibfield  {journal} {\bibinfo
  {journal} {arXiv preprint arXiv:1004.2515}\ } (\bibinfo {year}
  {2010})}\BibitemShut {NoStop}%
\bibitem [{\citenamefont {Wibral}\ \emph {et~al.}(2017)\citenamefont {Wibral},
  \citenamefont {Priesemann}, \citenamefont {Kay}, \citenamefont {Lizier},\
  and\ \citenamefont {Phillips}}]{wibral_partial_2017}%
  \BibitemOpen
  \bibfield  {author} {\bibinfo {author} {\bibfnamefont {M.}~\bibnamefont
  {Wibral}}, \bibinfo {author} {\bibfnamefont {V.}~\bibnamefont {Priesemann}},
  \bibinfo {author} {\bibfnamefont {J.~W.}\ \bibnamefont {Kay}}, \bibinfo
  {author} {\bibfnamefont {J.~T.}\ \bibnamefont {Lizier}},\ and\ \bibinfo
  {author} {\bibfnamefont {W.~A.}\ \bibnamefont {Phillips}},\ }\bibfield
  {title} {\bibinfo {title} {Partial information decomposition as a unified
  approach to the specification of neural goal functions},\ }\href@noop {}
  {\bibfield  {journal} {\bibinfo  {journal} {Brain and Cognition}\ }\bibinfo
  {series} {Perspectives on {Human} {Probabilistic} {Inferences} and the
  '{Bayesian} {Brain}'},\ \textbf {\bibinfo {volume} {112}},\ \bibinfo {pages}
  {25} (\bibinfo {year} {2017})}\BibitemShut {NoStop}%
\bibitem [{\citenamefont {Lizier}\ \emph {et~al.}(2018)\citenamefont {Lizier},
  \citenamefont {Bertschinger}, \citenamefont {Jost},\ and\ \citenamefont
  {Wibral}}]{lizier2018information}%
  \BibitemOpen
  \bibfield  {author} {\bibinfo {author} {\bibfnamefont {J.}~\bibnamefont
  {Lizier}}, \bibinfo {author} {\bibfnamefont {N.}~\bibnamefont
  {Bertschinger}}, \bibinfo {author} {\bibfnamefont {J.}~\bibnamefont {Jost}},\
  and\ \bibinfo {author} {\bibfnamefont {M.}~\bibnamefont {Wibral}},\
  }\href@noop {} {\bibinfo {title} {Information decomposition of target effects
  from multi-source interactions: perspectives on previous, current and future
  work}} (\bibinfo {year} {2018})\BibitemShut {NoStop}%
\bibitem [{\citenamefont
  {Kolchinsky}(2022)}]{kolchinskyNovelApproachPartial2022}%
  \BibitemOpen
  \bibfield  {author} {\bibinfo {author} {\bibfnamefont {A.}~\bibnamefont
  {Kolchinsky}},\ }\bibfield  {title} {\bibinfo {title} {A {{Novel Approach}}
  to the {{Partial Information Decomposition}}},\ }\href
  {https://doi.org/10.3390/e24030403} {\bibfield  {journal} {\bibinfo
  {journal} {Entropy}\ }\textbf {\bibinfo {volume} {24}},\ \bibinfo {pages}
  {403} (\bibinfo {year} {2022})}\BibitemShut {NoStop}%
\bibitem [{\citenamefont {Williams}(2011)}]{williams_information_2011}%
  \BibitemOpen
  \bibfield  {author} {\bibinfo {author} {\bibfnamefont {P.~L.}\ \bibnamefont
  {Williams}},\ }{\selectlanguage {English}\emph {\bibinfo {title} {Information
  dynamics: {Its} theory and application to embodied cognitive systems}}},\
  \href@noop {} {\bibinfo {type} {Ph.{D}.}} (\bibinfo {year}
  {2011})\BibitemShut {NoStop}%
\bibitem [{\citenamefont {Tishby}\ \emph {et~al.}(1999)\citenamefont {Tishby},
  \citenamefont {Pereira},\ and\ \citenamefont
  {Bialek}}]{tishbyInformationBottleneckMethod1999}%
  \BibitemOpen
  \bibfield  {author} {\bibinfo {author} {\bibfnamefont {N.}~\bibnamefont
  {Tishby}}, \bibinfo {author} {\bibfnamefont {F.}~\bibnamefont {Pereira}},\
  and\ \bibinfo {author} {\bibfnamefont {W.}~\bibnamefont {Bialek}},\
  }\bibfield  {title} {\bibinfo {title} {The information bottleneck method},\
  }in\ \href@noop {} {\emph {\bibinfo {booktitle} {37th {Allerton} {Conf} on
  {Communication}}}}\ (\bibinfo {year} {1999})\BibitemShut {NoStop}%
\bibitem [{\citenamefont {Hu}\ \emph {et~al.}(2024)\citenamefont {Hu},
  \citenamefont {Lou}, \citenamefont {Yan},\ and\ \citenamefont
  {Ye}}]{hu2024survey}%
  \BibitemOpen
  \bibfield  {author} {\bibinfo {author} {\bibfnamefont {S.}~\bibnamefont
  {Hu}}, \bibinfo {author} {\bibfnamefont {Z.}~\bibnamefont {Lou}}, \bibinfo
  {author} {\bibfnamefont {X.}~\bibnamefont {Yan}},\ and\ \bibinfo {author}
  {\bibfnamefont {Y.}~\bibnamefont {Ye}},\ }\bibfield  {title} {\bibinfo
  {title} {A survey on information bottleneck},\ }\href@noop {} {\bibfield
  {journal} {\bibinfo  {journal} {IEEE Transactions on Pattern Analysis and
  Machine Intelligence}\ } (\bibinfo {year} {2024})}\BibitemShut {NoStop}%
\bibitem [{\citenamefont {Palmer}\ \emph {et~al.}(2015)\citenamefont {Palmer},
  \citenamefont {Marre}, \citenamefont {Berry},\ and\ \citenamefont
  {Bialek}}]{palmer2015predictive}%
  \BibitemOpen
  \bibfield  {author} {\bibinfo {author} {\bibfnamefont {S.~E.}\ \bibnamefont
  {Palmer}}, \bibinfo {author} {\bibfnamefont {O.}~\bibnamefont {Marre}},
  \bibinfo {author} {\bibfnamefont {M.~J.}\ \bibnamefont {Berry}},\ and\
  \bibinfo {author} {\bibfnamefont {W.}~\bibnamefont {Bialek}},\ }\bibfield
  {title} {\bibinfo {title} {Predictive information in a sensory population},\
  }\href@noop {} {\bibfield  {journal} {\bibinfo  {journal} {Proceedings of the
  National Academy of Sciences}\ }\textbf {\bibinfo {volume} {112}},\ \bibinfo
  {pages} {6908} (\bibinfo {year} {2015})}\BibitemShut {NoStop}%
\bibitem [{\citenamefont {Wang}\ \emph {et~al.}(2019)\citenamefont {Wang},
  \citenamefont {Ribeiro},\ and\ \citenamefont
  {Tiwary}}]{wangFutureInformationBottleneck2019}%
  \BibitemOpen
  \bibfield  {author} {\bibinfo {author} {\bibfnamefont {Y.}~\bibnamefont
  {Wang}}, \bibinfo {author} {\bibfnamefont {J.~M.~L.}\ \bibnamefont
  {Ribeiro}},\ and\ \bibinfo {author} {\bibfnamefont {P.}~\bibnamefont
  {Tiwary}},\ }\bibfield  {title} {\bibinfo {title} {Past–future information
  bottleneck for sampling molecular reaction coordinate simultaneously with
  thermodynamics and kinetics},\ }\href
  {https://doi.org/10.1038/s41467-019-11405-4} {\bibfield  {journal} {\bibinfo
  {journal} {Nature Communications}\ }\textbf {\bibinfo {volume} {10}},\
  \bibinfo {pages} {3573} (\bibinfo {year} {2019})},\ \bibinfo {note}
  {publisher: Nature Publishing Group}\BibitemShut {NoStop}%
\bibitem [{\citenamefont {Zaslavsky}\ \emph {et~al.}(2018)\citenamefont
  {Zaslavsky}, \citenamefont {Kemp}, \citenamefont {Regier},\ and\
  \citenamefont {Tishby}}]{zaslavskyEfficientCompressionColor2018}%
  \BibitemOpen
  \bibfield  {author} {\bibinfo {author} {\bibfnamefont {N.}~\bibnamefont
  {Zaslavsky}}, \bibinfo {author} {\bibfnamefont {C.}~\bibnamefont {Kemp}},
  \bibinfo {author} {\bibfnamefont {T.}~\bibnamefont {Regier}},\ and\ \bibinfo
  {author} {\bibfnamefont {N.}~\bibnamefont {Tishby}},\ }\bibfield  {title}
  {\bibinfo {title} {Efficient compression in color naming and its evolution},\
  }\href {https://doi.org/10.1073/pnas.1800521115} {\bibfield  {journal}
  {\bibinfo  {journal} {Proceedings of the National Academy of Sciences}\
  }\textbf {\bibinfo {volume} {115}},\ \bibinfo {pages} {7937} (\bibinfo {year}
  {2018})}\BibitemShut {NoStop}%
\bibitem [{\citenamefont {Alemi}\ \emph {et~al.}(2017)\citenamefont {Alemi},
  \citenamefont {Fischer}, \citenamefont {Dillon},\ and\ \citenamefont
  {Murphy}}]{alemi2016deep}%
  \BibitemOpen
  \bibfield  {author} {\bibinfo {author} {\bibfnamefont {A.~A.}\ \bibnamefont
  {Alemi}}, \bibinfo {author} {\bibfnamefont {I.}~\bibnamefont {Fischer}},
  \bibinfo {author} {\bibfnamefont {J.~V.}\ \bibnamefont {Dillon}},\ and\
  \bibinfo {author} {\bibfnamefont {K.}~\bibnamefont {Murphy}},\ }\bibfield
  {title} {\bibinfo {title} {Deep variational information bottleneck},\
  }\href@noop {} {\bibfield  {journal} {\bibinfo  {journal} {ICLR}\ } (\bibinfo
  {year} {2017})}\BibitemShut {NoStop}%
\bibitem [{\citenamefont {Kolchinsky}\ \emph
  {et~al.}(2019{\natexlab{a}})\citenamefont {Kolchinsky}, \citenamefont
  {Tracey},\ and\ \citenamefont {Wolpert}}]{kolchinsky2019nonlinear}%
  \BibitemOpen
  \bibfield  {author} {\bibinfo {author} {\bibfnamefont {A.}~\bibnamefont
  {Kolchinsky}}, \bibinfo {author} {\bibfnamefont {B.~D.}\ \bibnamefont
  {Tracey}},\ and\ \bibinfo {author} {\bibfnamefont {D.~H.}\ \bibnamefont
  {Wolpert}},\ }\bibfield  {title} {\bibinfo {title} {Nonlinear information
  bottleneck},\ }\href@noop {} {\bibfield  {journal} {\bibinfo  {journal}
  {Entropy}\ }\textbf {\bibinfo {volume} {21}},\ \bibinfo {pages} {1181}
  (\bibinfo {year} {2019}{\natexlab{a}})}\BibitemShut {NoStop}%
\bibitem [{\citenamefont {Fischer}(2020)}]{fischer2020conditional}%
  \BibitemOpen
  \bibfield  {author} {\bibinfo {author} {\bibfnamefont {I.}~\bibnamefont
  {Fischer}},\ }\bibfield  {title} {\bibinfo {title} {The conditional entropy
  bottleneck},\ }\href@noop {} {\bibfield  {journal} {\bibinfo  {journal}
  {Entropy}\ }\textbf {\bibinfo {volume} {22}},\ \bibinfo {pages} {999}
  (\bibinfo {year} {2020})}\BibitemShut {NoStop}%
\bibitem [{\citenamefont {Goldfeld}\ and\ \citenamefont
  {Polyanskiy}(2020)}]{goldfeld2020information}%
  \BibitemOpen
  \bibfield  {author} {\bibinfo {author} {\bibfnamefont {Z.}~\bibnamefont
  {Goldfeld}}\ and\ \bibinfo {author} {\bibfnamefont {Y.}~\bibnamefont
  {Polyanskiy}},\ }\bibfield  {title} {\bibinfo {title} {The information
  bottleneck problem and its applications in machine learning},\ }\href@noop {}
  {\bibfield  {journal} {\bibinfo  {journal} {IEEE Journal on Selected Areas in
  Information Theory}\ }\textbf {\bibinfo {volume} {1}},\ \bibinfo {pages} {19}
  (\bibinfo {year} {2020})}\BibitemShut {NoStop}%
\bibitem [{\citenamefont {Ahlswede}\ and\ \citenamefont
  {Körner}(1975)}]{ahlswedeSourceCodingSide1975}%
  \BibitemOpen
  \bibfield  {author} {\bibinfo {author} {\bibfnamefont {R.}~\bibnamefont
  {Ahlswede}}\ and\ \bibinfo {author} {\bibfnamefont {J.}~\bibnamefont
  {Körner}},\ }\bibfield  {title} {\bibinfo {title} {Source {Coding} with
  {Side} {Information} and a {Converse} for {Degraded} {Broadcast}
  {Channels}},\ }\href {https://doi.org/10.1109/TIT.1975.1055469} {\bibfield
  {journal} {\bibinfo  {journal} {IEEE Transactions on Information Theory}\ ,\
  \bibinfo {pages} {9}} (\bibinfo {year} {1975})}\BibitemShut {NoStop}%
\bibitem [{\citenamefont {Witsenhausen}\ and\ \citenamefont
  {Wyner}(1975)}]{witsenhausen_conditional_1975}%
  \BibitemOpen
  \bibfield  {author} {\bibinfo {author} {\bibfnamefont {H.}~\bibnamefont
  {Witsenhausen}}\ and\ \bibinfo {author} {\bibfnamefont {A.}~\bibnamefont
  {Wyner}},\ }\bibfield  {title} {\bibinfo {title} {A conditional entropy bound
  for a pair of discrete random variables},\ }\href
  {https://doi.org/10.1109/TIT.1975.1055437} {\bibfield  {journal} {\bibinfo
  {journal} {IEEE Transactions on Information Theory}\ }\textbf {\bibinfo
  {volume} {21}},\ \bibinfo {pages} {493} (\bibinfo {year} {1975})}\BibitemShut
  {NoStop}%
\bibitem [{\citenamefont {Gilad-Bachrach}\ \emph {et~al.}(2003)\citenamefont
  {Gilad-Bachrach}, \citenamefont {Navot},\ and\ \citenamefont
  {Tishby}}]{goos_information_2003}%
  \BibitemOpen
  \bibfield  {author} {\bibinfo {author} {\bibfnamefont {R.}~\bibnamefont
  {Gilad-Bachrach}}, \bibinfo {author} {\bibfnamefont {A.}~\bibnamefont
  {Navot}},\ and\ \bibinfo {author} {\bibfnamefont {N.}~\bibnamefont
  {Tishby}},\ }\bibfield  {title} {\bibinfo {title} {An {Information}
  {Theoretic} {Tradeoff} between {Complexity} and {Accuracy}},\ }in\ \href
  {https://doi.org/10.1007/978-3-540-45167-9_43} {\emph {\bibinfo {booktitle}
  {Learning {Theory} and {Kernel} {Machines}}}},\ Vol.\ \bibinfo {volume}
  {2777},\ \bibinfo {editor} {edited by\ \bibinfo {editor} {\bibfnamefont
  {G.}~\bibnamefont {Goos}}, \bibinfo {editor} {\bibfnamefont {J.}~\bibnamefont
  {Hartmanis}}, \bibinfo {editor} {\bibfnamefont {J.}~\bibnamefont {van
  Leeuwen}}, \bibinfo {editor} {\bibfnamefont {B.}~\bibnamefont {Schölkopf}},\
  and\ \bibinfo {editor} {\bibfnamefont {M.~K.}\ \bibnamefont {Warmuth}}}\
  (\bibinfo  {publisher} {Springer Berlin Heidelberg},\ \bibinfo {address}
  {Berlin, Heidelberg},\ \bibinfo {year} {2003})\ pp.\ \bibinfo {pages}
  {595--609}\BibitemShut {NoStop}%
\bibitem [{\citenamefont {Kolchinsky}\ \emph
  {et~al.}(2019{\natexlab{b}})\citenamefont {Kolchinsky}, \citenamefont
  {Tracey},\ and\ \citenamefont
  {Van~Kuyk}}]{kolchinskyCaveatsInformationBottleneck2019}%
  \BibitemOpen
  \bibfield  {author} {\bibinfo {author} {\bibfnamefont {A.}~\bibnamefont
  {Kolchinsky}}, \bibinfo {author} {\bibfnamefont {B.~D.}\ \bibnamefont
  {Tracey}},\ and\ \bibinfo {author} {\bibfnamefont {S.}~\bibnamefont
  {Van~Kuyk}},\ }\bibfield  {title} {\bibinfo {title} {Caveats for information
  bottleneck in deterministic scenarios},\ }in\ \href
  {http://arxiv.org/abs/1808.07593} {\emph {\bibinfo {booktitle} {{ICLR}
  2019}}}\ (\bibinfo {year} {2019})\ \bibinfo {note} {arXiv:
  1808.07593}\BibitemShut {NoStop}%
\bibitem [{\citenamefont {Rodr{\'\i}guez~G{\'a}lvez}\ \emph
  {et~al.}(2020)\citenamefont {Rodr{\'\i}guez~G{\'a}lvez}, \citenamefont
  {Thobaben},\ and\ \citenamefont {Skoglund}}]{rodriguez2020convex}%
  \BibitemOpen
  \bibfield  {author} {\bibinfo {author} {\bibfnamefont {B.}~\bibnamefont
  {Rodr{\'\i}guez~G{\'a}lvez}}, \bibinfo {author} {\bibfnamefont
  {R.}~\bibnamefont {Thobaben}},\ and\ \bibinfo {author} {\bibfnamefont
  {M.}~\bibnamefont {Skoglund}},\ }\bibfield  {title} {\bibinfo {title} {The
  convex information bottleneck lagrangian},\ }\href@noop {} {\bibfield
  {journal} {\bibinfo  {journal} {Entropy}\ }\textbf {\bibinfo {volume} {22}},\
  \bibinfo {pages} {98} (\bibinfo {year} {2020})}\BibitemShut {NoStop}%
\bibitem [{\citenamefont {Benger}\ \emph {et~al.}(2023)\citenamefont {Benger},
  \citenamefont {Asoodeh},\ and\ \citenamefont {Chen}}]{benger2023cardinality}%
  \BibitemOpen
  \bibfield  {author} {\bibinfo {author} {\bibfnamefont {E.}~\bibnamefont
  {Benger}}, \bibinfo {author} {\bibfnamefont {S.}~\bibnamefont {Asoodeh}},\
  and\ \bibinfo {author} {\bibfnamefont {J.}~\bibnamefont {Chen}},\ }\bibfield
  {title} {\bibinfo {title} {The cardinality bound on the information
  bottleneck representations is tight},\ }in\ \href@noop {} {\emph {\bibinfo
  {booktitle} {2023 IEEE International Symposium on Information Theory
  (ISIT)}}}\ (\bibinfo {organization} {IEEE},\ \bibinfo {year} {2023})\ pp.\
  \bibinfo {pages} {1478--1483}\BibitemShut {NoStop}%
\bibitem [{\citenamefont {Geiger}\ and\ \citenamefont
  {Fischer}(2020)}]{geiger2020comparison}%
  \BibitemOpen
  \bibfield  {author} {\bibinfo {author} {\bibfnamefont {B.~C.}\ \bibnamefont
  {Geiger}}\ and\ \bibinfo {author} {\bibfnamefont {I.~S.}\ \bibnamefont
  {Fischer}},\ }\bibfield  {title} {\bibinfo {title} {A comparison of
  variational bounds for the information bottleneck functional},\ }\href@noop
  {} {\bibfield  {journal} {\bibinfo  {journal} {Entropy}\ }\textbf {\bibinfo
  {volume} {22}},\ \bibinfo {pages} {1229} (\bibinfo {year}
  {2020})}\BibitemShut {NoStop}%
\bibitem [{\citenamefont {Federici}\ \emph {et~al.}(2020)\citenamefont
  {Federici}, \citenamefont {Dutta}, \citenamefont {Forr{\'e}}, \citenamefont
  {Kushman},\ and\ \citenamefont {Akata}}]{federici2020learning}%
  \BibitemOpen
  \bibfield  {author} {\bibinfo {author} {\bibfnamefont {M.}~\bibnamefont
  {Federici}}, \bibinfo {author} {\bibfnamefont {A.}~\bibnamefont {Dutta}},
  \bibinfo {author} {\bibfnamefont {P.}~\bibnamefont {Forr{\'e}}}, \bibinfo
  {author} {\bibfnamefont {N.}~\bibnamefont {Kushman}},\ and\ \bibinfo {author}
  {\bibfnamefont {Z.}~\bibnamefont {Akata}},\ }\bibfield  {title} {\bibinfo
  {title} {Learning robust representations via multi-view information
  bottleneck},\ }\href@noop {} {\bibfield  {journal} {\bibinfo  {journal}
  {ICLR}\ } (\bibinfo {year} {2020})}\BibitemShut {NoStop}%
\bibitem [{\citenamefont {Murphy}\ and\ \citenamefont
  {Bassett}(2024)}]{murphy2024machine}%
  \BibitemOpen
  \bibfield  {author} {\bibinfo {author} {\bibfnamefont {K.~A.}\ \bibnamefont
  {Murphy}}\ and\ \bibinfo {author} {\bibfnamefont {D.~S.}\ \bibnamefont
  {Bassett}},\ }\bibfield  {title} {\bibinfo {title} {Machine-learning
  optimized measurements of chaotic dynamical systems via the information
  bottleneck},\ }\href@noop {} {\bibfield  {journal} {\bibinfo  {journal}
  {Physical Review Letters}\ }\textbf {\bibinfo {volume} {132}},\ \bibinfo
  {pages} {197201} (\bibinfo {year} {2024})}\BibitemShut {NoStop}%
\bibitem [{\citenamefont {Slonim}\ \emph {et~al.}(2006)\citenamefont {Slonim},
  \citenamefont {Friedman},\ and\ \citenamefont
  {Tishby}}]{slonimMultivariateInformationBottleneck2006}%
  \BibitemOpen
  \bibfield  {author} {\bibinfo {author} {\bibfnamefont {N.}~\bibnamefont
  {Slonim}}, \bibinfo {author} {\bibfnamefont {N.}~\bibnamefont {Friedman}},\
  and\ \bibinfo {author} {\bibfnamefont {N.}~\bibnamefont {Tishby}},\
  }\bibfield  {title} {\bibinfo {title} {Multivariate {Information}
  {Bottleneck}},\ }\bibfield  {journal} {\bibinfo  {journal} {Neural
  Computation}\ }\textbf {\bibinfo {volume} {18}},\ \href
  {https://doi.org/10.1162/neco.2006.18.8.1739} {10.1162/neco.2006.18.8.1739}
  (\bibinfo {year} {2006})\BibitemShut {NoStop}%
\bibitem [{\citenamefont {Shannon}(1953)}]{shannon_lattice_1953}%
  \BibitemOpen
  \bibfield  {author} {\bibinfo {author} {\bibfnamefont {C.}~\bibnamefont
  {Shannon}},\ }\bibfield  {title} {\bibinfo {title} {The lattice theory of
  information},\ }\href@noop {} {\bibfield  {journal} {\bibinfo  {journal}
  {Transactions of the IRE Professional Group on Information Theory}\ }\textbf
  {\bibinfo {volume} {1}},\ \bibinfo {pages} {105} (\bibinfo {year}
  {1953})}\BibitemShut {NoStop}%
\bibitem [{\citenamefont {McGill}(1954)}]{mcgill1954multivariate}%
  \BibitemOpen
  \bibfield  {author} {\bibinfo {author} {\bibfnamefont {W.}~\bibnamefont
  {McGill}},\ }\bibfield  {title} {\bibinfo {title} {Multivariate information
  transmission},\ }\href@noop {} {\bibfield  {journal} {\bibinfo  {journal}
  {Transactions of the IRE Professional Group on Information Theory}\ }\textbf
  {\bibinfo {volume} {4}},\ \bibinfo {pages} {93} (\bibinfo {year}
  {1954})}\BibitemShut {NoStop}%
\bibitem [{\citenamefont {Reza}(1961)}]{rezaIntroductionInformationTheory1961}%
  \BibitemOpen
  \bibfield  {author} {\bibinfo {author} {\bibfnamefont {F.~M.}\ \bibnamefont
  {Reza}},\ }\href@noop {} {\emph {\bibinfo {title} {An Introduction to
  Information Theory}}}\ (\bibinfo  {publisher} {{Dover Publications, Inc}},\
  \bibinfo {year} {1961})\BibitemShut {NoStop}%
\bibitem [{\citenamefont {Ting}(1962)}]{ting1962amount}%
  \BibitemOpen
  \bibfield  {author} {\bibinfo {author} {\bibfnamefont {H.~K.}\ \bibnamefont
  {Ting}},\ }\bibfield  {title} {\bibinfo {title} {On the amount of
  information},\ }\href@noop {} {\bibfield  {journal} {\bibinfo  {journal}
  {Theory of Probability \& Its Applications}\ }\textbf {\bibinfo {volume}
  {7}},\ \bibinfo {pages} {439} (\bibinfo {year} {1962})}\BibitemShut {NoStop}%
\bibitem [{\citenamefont {Han}(1975)}]{han1975linear}%
  \BibitemOpen
  \bibfield  {author} {\bibinfo {author} {\bibfnamefont {T.}~\bibnamefont
  {Han}},\ }\bibfield  {title} {\bibinfo {title} {Linear dependence structure
  of the entropy space},\ }\href@noop {} {\bibfield  {journal} {\bibinfo
  {journal} {Information and Control}\ }\textbf {\bibinfo {volume} {29}},\
  \bibinfo {pages} {337} (\bibinfo {year} {1975})}\BibitemShut {NoStop}%
\bibitem [{\citenamefont {Yeung}(1991)}]{yeung1991new}%
  \BibitemOpen
  \bibfield  {author} {\bibinfo {author} {\bibfnamefont {R.~W.}\ \bibnamefont
  {Yeung}},\ }\bibfield  {title} {\bibinfo {title} {A new outlook on shannon's
  information measures},\ }\href@noop {} {\bibfield  {journal} {\bibinfo
  {journal} {IEEE transactions on information theory}\ }\textbf {\bibinfo
  {volume} {37}},\ \bibinfo {pages} {466} (\bibinfo {year} {1991})}\BibitemShut
  {NoStop}%
\bibitem [{\citenamefont {Bell}(2003)}]{bell2003co}%
  \BibitemOpen
  \bibfield  {author} {\bibinfo {author} {\bibfnamefont {A.~J.}\ \bibnamefont
  {Bell}},\ }\bibfield  {title} {\bibinfo {title} {The co-information
  lattice},\ }in\ \href@noop {} {\emph {\bibinfo {booktitle} {Proceedings of
  the Fifth International Workshop on Independent Component Analysis and Blind
  Signal Separation: ICA}}},\ Vol.\ \bibinfo {volume} {2003}\ (\bibinfo {year}
  {2003})\BibitemShut {NoStop}%
\bibitem [{\citenamefont {Gomes}\ and\ \citenamefont
  {Figueiredo}(2023)}]{gomes2023orders}%
  \BibitemOpen
  \bibfield  {author} {\bibinfo {author} {\bibfnamefont {A.~F.}\ \bibnamefont
  {Gomes}}\ and\ \bibinfo {author} {\bibfnamefont {M.~A.}\ \bibnamefont
  {Figueiredo}},\ }\bibfield  {title} {\bibinfo {title} {Orders between
  channels and implications for partial information decomposition},\
  }\href@noop {} {\bibfield  {journal} {\bibinfo  {journal} {Entropy}\ }\textbf
  {\bibinfo {volume} {25}},\ \bibinfo {pages} {975} (\bibinfo {year}
  {2023})}\BibitemShut {NoStop}%
\bibitem [{\citenamefont {Griffith}\ and\ \citenamefont
  {Koch}(2014)}]{griffith2014quantifying}%
  \BibitemOpen
  \bibfield  {author} {\bibinfo {author} {\bibfnamefont {V.}~\bibnamefont
  {Griffith}}\ and\ \bibinfo {author} {\bibfnamefont {C.}~\bibnamefont
  {Koch}},\ }\bibfield  {title} {\bibinfo {title} {Quantifying synergistic
  mutual information},\ }in\ \href@noop {} {\emph {\bibinfo {booktitle} {Guided
  Self-Organization: Inception}}}\ (\bibinfo  {publisher} {Springer},\ \bibinfo
  {year} {2014})\ pp.\ \bibinfo {pages} {159--190}\BibitemShut {NoStop}%
\bibitem [{\citenamefont {Griffith}\ \emph {et~al.}(2014)\citenamefont
  {Griffith}, \citenamefont {Chong}, \citenamefont {James}, \citenamefont
  {Ellison},\ and\ \citenamefont {Crutchfield}}]{griffith2014intersection}%
  \BibitemOpen
  \bibfield  {author} {\bibinfo {author} {\bibfnamefont {V.}~\bibnamefont
  {Griffith}}, \bibinfo {author} {\bibfnamefont {E.~K.}\ \bibnamefont {Chong}},
  \bibinfo {author} {\bibfnamefont {R.~G.}\ \bibnamefont {James}}, \bibinfo
  {author} {\bibfnamefont {C.~J.}\ \bibnamefont {Ellison}},\ and\ \bibinfo
  {author} {\bibfnamefont {J.~P.}\ \bibnamefont {Crutchfield}},\ }\bibfield
  {title} {\bibinfo {title} {Intersection information based on common
  randomness},\ }\href@noop {} {\bibfield  {journal} {\bibinfo  {journal}
  {Entropy}\ }\textbf {\bibinfo {volume} {16}},\ \bibinfo {pages} {1985}
  (\bibinfo {year} {2014})}\BibitemShut {NoStop}%
\bibitem [{\citenamefont {Griffith}\ and\ \citenamefont
  {Ho}(2015)}]{griffith_quantifying_2015}%
  \BibitemOpen
  \bibfield  {author} {\bibinfo {author} {\bibfnamefont {V.}~\bibnamefont
  {Griffith}}\ and\ \bibinfo {author} {\bibfnamefont {T.}~\bibnamefont {Ho}},\
  }\bibfield  {title} {\bibinfo {title} {Quantifying redundant information in
  predicting a target random variable},\ }\href@noop {} {\bibfield  {journal}
  {\bibinfo  {journal} {Entropy}\ }\textbf {\bibinfo {volume} {17}},\ \bibinfo
  {pages} {4644} (\bibinfo {year} {2015})}\BibitemShut {NoStop}%
\bibitem [{\citenamefont {Bertschinger}\ and\ \citenamefont
  {Rauh}(2014)}]{bertschinger2014blackwell}%
  \BibitemOpen
  \bibfield  {author} {\bibinfo {author} {\bibfnamefont {N.}~\bibnamefont
  {Bertschinger}}\ and\ \bibinfo {author} {\bibfnamefont {J.}~\bibnamefont
  {Rauh}},\ }\bibfield  {title} {\bibinfo {title} {The blackwell relation
  defines no lattice},\ }in\ \href@noop {} {\emph {\bibinfo {booktitle} {2014
  IEEE International Symposium on Information Theory}}}\ (\bibinfo
  {organization} {IEEE},\ \bibinfo {year} {2014})\ pp.\ \bibinfo {pages}
  {2479--2483}\BibitemShut {NoStop}%
\bibitem [{\citenamefont {Blackwell}(1953)}]{blackwell_equivalent_1953}%
  \BibitemOpen
  \bibfield  {author} {\bibinfo {author} {\bibfnamefont {D.}~\bibnamefont
  {Blackwell}},\ }\bibfield  {title} {\bibinfo {title} {Equivalent comparisons
  of experiments},\ }\href@noop {} {\bibfield  {journal} {\bibinfo  {journal}
  {The annals of mathematical statistics}\ ,\ \bibinfo {pages} {265}} (\bibinfo
  {year} {1953})}\BibitemShut {NoStop}%
\bibitem [{\citenamefont {Rauh}\ \emph
  {et~al.}(2017{\natexlab{a}})\citenamefont {Rauh}, \citenamefont {Banerjee},
  \citenamefont {Olbrich}, \citenamefont {Jost}, \citenamefont {Bertschinger},\
  and\ \citenamefont {Wolpert}}]{rauh_coarse-graining_2017}%
  \BibitemOpen
  \bibfield  {author} {\bibinfo {author} {\bibfnamefont {J.}~\bibnamefont
  {Rauh}}, \bibinfo {author} {\bibfnamefont {P.~K.}\ \bibnamefont {Banerjee}},
  \bibinfo {author} {\bibfnamefont {E.}~\bibnamefont {Olbrich}}, \bibinfo
  {author} {\bibfnamefont {J.}~\bibnamefont {Jost}}, \bibinfo {author}
  {\bibfnamefont {N.}~\bibnamefont {Bertschinger}},\ and\ \bibinfo {author}
  {\bibfnamefont {D.}~\bibnamefont {Wolpert}},\ }\bibfield  {title} {\bibinfo
  {title} {Coarse-{Graining} and the {Blackwell} {Order}},\ }\href@noop {}
  {\bibfield  {journal} {\bibinfo  {journal} {Entropy}\ }\textbf {\bibinfo
  {volume} {19}},\ \bibinfo {pages} {527} (\bibinfo {year}
  {2017}{\natexlab{a}})}\BibitemShut {NoStop}%
\bibitem [{\citenamefont {Bertschinger}\ \emph {et~al.}(2014)\citenamefont
  {Bertschinger}, \citenamefont {Rauh}, \citenamefont {Olbrich}, \citenamefont
  {Jost},\ and\ \citenamefont {Ay}}]{bertschinger2014quantifying}%
  \BibitemOpen
  \bibfield  {author} {\bibinfo {author} {\bibfnamefont {N.}~\bibnamefont
  {Bertschinger}}, \bibinfo {author} {\bibfnamefont {J.}~\bibnamefont {Rauh}},
  \bibinfo {author} {\bibfnamefont {E.}~\bibnamefont {Olbrich}}, \bibinfo
  {author} {\bibfnamefont {J.}~\bibnamefont {Jost}},\ and\ \bibinfo {author}
  {\bibfnamefont {N.}~\bibnamefont {Ay}},\ }\bibfield  {title} {\bibinfo
  {title} {Quantifying unique information},\ }\href@noop {} {\bibfield
  {journal} {\bibinfo  {journal} {Entropy}\ }\textbf {\bibinfo {volume} {16}},\
  \bibinfo {pages} {2161} (\bibinfo {year} {2014})}\BibitemShut {NoStop}%
\bibitem [{\citenamefont {Rauh}\ \emph
  {et~al.}(2017{\natexlab{b}})\citenamefont {Rauh}, \citenamefont {Banerjee},
  \citenamefont {Olbrich}, \citenamefont {Jost},\ and\ \citenamefont
  {Bertschinger}}]{rauh2017extractable}%
  \BibitemOpen
  \bibfield  {author} {\bibinfo {author} {\bibfnamefont {J.}~\bibnamefont
  {Rauh}}, \bibinfo {author} {\bibfnamefont {P.~K.}\ \bibnamefont {Banerjee}},
  \bibinfo {author} {\bibfnamefont {E.}~\bibnamefont {Olbrich}}, \bibinfo
  {author} {\bibfnamefont {J.}~\bibnamefont {Jost}},\ and\ \bibinfo {author}
  {\bibfnamefont {N.}~\bibnamefont {Bertschinger}},\ }\bibfield  {title}
  {\bibinfo {title} {On extractable shared information},\ }\href@noop {}
  {\bibfield  {journal} {\bibinfo  {journal} {Entropy}\ }\textbf {\bibinfo
  {volume} {19}},\ \bibinfo {pages} {328} (\bibinfo {year}
  {2017}{\natexlab{b}})}\BibitemShut {NoStop}%
\bibitem [{\citenamefont {Venkatesh}\ and\ \citenamefont
  {Schamberg}(2022)}]{venkatesh2022partial}%
  \BibitemOpen
  \bibfield  {author} {\bibinfo {author} {\bibfnamefont {P.}~\bibnamefont
  {Venkatesh}}\ and\ \bibinfo {author} {\bibfnamefont {G.}~\bibnamefont
  {Schamberg}},\ }\bibfield  {title} {\bibinfo {title} {Partial information
  decomposition via deficiency for multivariate gaussians},\ }in\ \href@noop {}
  {\emph {\bibinfo {booktitle} {2022 IEEE International Symposium on
  Information Theory (ISIT)}}}\ (\bibinfo {organization} {IEEE},\ \bibinfo
  {year} {2022})\ pp.\ \bibinfo {pages} {2892--2897}\BibitemShut {NoStop}%
\bibitem [{\citenamefont {Mages}\ \emph {et~al.}(2024)\citenamefont {Mages},
  \citenamefont {Anastasiadi},\ and\ \citenamefont {Rohner}}]{mages2024non}%
  \BibitemOpen
  \bibfield  {author} {\bibinfo {author} {\bibfnamefont {T.}~\bibnamefont
  {Mages}}, \bibinfo {author} {\bibfnamefont {E.}~\bibnamefont {Anastasiadi}},\
  and\ \bibinfo {author} {\bibfnamefont {C.}~\bibnamefont {Rohner}},\
  }\bibfield  {title} {\bibinfo {title} {Non-negative decomposition of
  multivariate information: From minimum to blackwell specific information},\
  }\href@noop {} {\bibfield  {journal} {\bibinfo  {journal}
  {10.20944/preprints202403.0285.v2}\ } (\bibinfo {year} {2024})}\BibitemShut
  {NoStop}%
\bibitem [{\citenamefont {Le~Cam}(1964)}]{le1964sufficiency}%
  \BibitemOpen
  \bibfield  {author} {\bibinfo {author} {\bibfnamefont {L.}~\bibnamefont
  {Le~Cam}},\ }\bibfield  {title} {\bibinfo {title} {Sufficiency and
  approximate sufficiency},\ }\href@noop {} {\bibfield  {journal} {\bibinfo
  {journal} {The Annals of Mathematical Statistics}\ ,\ \bibinfo {pages}
  {1419}} (\bibinfo {year} {1964})}\BibitemShut {NoStop}%
\bibitem [{\citenamefont {Raginsky}(2011)}]{raginsky_shannon_2011}%
  \BibitemOpen
  \bibfield  {author} {\bibinfo {author} {\bibfnamefont {M.}~\bibnamefont
  {Raginsky}},\ }\bibfield  {title} {\bibinfo {title} {Shannon meets
  {Blackwell} and {Le} {Cam}: {Channels}, codes, and statistical experiments}\
  }(\bibinfo  {publisher} {IEEE},\ \bibinfo {year} {2011})\ pp.\ \bibinfo
  {pages} {1220--1224}\BibitemShut {NoStop}%
\bibitem [{\citenamefont {Banerjee}\ \emph {et~al.}(2018)\citenamefont
  {Banerjee}, \citenamefont {Olbrich}, \citenamefont {Jost},\ and\
  \citenamefont {Rauh}}]{banerjee_unique_2018}%
  \BibitemOpen
  \bibfield  {author} {\bibinfo {author} {\bibfnamefont {P.~K.}\ \bibnamefont
  {Banerjee}}, \bibinfo {author} {\bibfnamefont {E.}~\bibnamefont {Olbrich}},
  \bibinfo {author} {\bibfnamefont {J.}~\bibnamefont {Jost}},\ and\ \bibinfo
  {author} {\bibfnamefont {J.}~\bibnamefont {Rauh}},\ }\bibfield  {title}
  {\bibinfo {title} {Unique informations and deficiencies},\ }in\ \href@noop {}
  {\emph {\bibinfo {booktitle} {2018 56th Annual Allerton Conference on
  Communication, Control, and Computing (Allerton)}}}\ (\bibinfo {organization}
  {IEEE},\ \bibinfo {year} {2018})\ pp.\ \bibinfo {pages} {32--38}\BibitemShut
  {NoStop}%
\bibitem [{\citenamefont {Banerjee}\ and\ \citenamefont
  {Montufar}(2020)}]{banerjeeVariationalDeficiencyBottleneck2020}%
  \BibitemOpen
  \bibfield  {author} {\bibinfo {author} {\bibfnamefont {P.~K.}\ \bibnamefont
  {Banerjee}}\ and\ \bibinfo {author} {\bibfnamefont {G.}~\bibnamefont
  {Montufar}},\ }\bibfield  {title} {\bibinfo {title} {The {Variational}
  {Deficiency} {Bottleneck}},\ }in\ \href
  {https://doi.org/10.1109/IJCNN48605.2020.9206900} {\emph {\bibinfo
  {booktitle} {2020 {International} {Joint} {Conference} on {Neural} {Networks}
  ({IJCNN})}}}\ (\bibinfo  {publisher} {IEEE},\ \bibinfo {address} {Glasgow,
  United Kingdom},\ \bibinfo {year} {2020})\ pp.\ \bibinfo {pages}
  {1--8}\BibitemShut {NoStop}%
\bibitem [{\citenamefont {Venkatesh}\ \emph {et~al.}(2023)\citenamefont
  {Venkatesh}, \citenamefont {Gurushankar},\ and\ \citenamefont
  {Schamberg}}]{venkateshCapturingInterpretingUnique2023}%
  \BibitemOpen
  \bibfield  {author} {\bibinfo {author} {\bibfnamefont {P.}~\bibnamefont
  {Venkatesh}}, \bibinfo {author} {\bibfnamefont {K.}~\bibnamefont
  {Gurushankar}},\ and\ \bibinfo {author} {\bibfnamefont {G.}~\bibnamefont
  {Schamberg}},\ }\bibfield  {title} {\bibinfo {title} {Capturing and
  {Interpreting} {Unique} {Information}},\ }in\ \href
  {https://doi.org/10.1109/ISIT54713.2023.10206597} {\emph {\bibinfo
  {booktitle} {2023 {IEEE} {International} {Symposium} on {Information}
  {Theory} ({ISIT})}}}\ (\bibinfo  {publisher} {IEEE},\ \bibinfo {address}
  {Taipei, Taiwan},\ \bibinfo {year} {2023})\ pp.\ \bibinfo {pages}
  {2631--2636}\BibitemShut {NoStop}%
\bibitem [{\citenamefont {Csiszár}\ and\ \citenamefont
  {Matus}(2003)}]{csiszarInformationProjectionsRevisited2003}%
  \BibitemOpen
  \bibfield  {author} {\bibinfo {author} {\bibfnamefont {I.}~\bibnamefont
  {Csiszár}}\ and\ \bibinfo {author} {\bibfnamefont {F.}~\bibnamefont
  {Matus}},\ }\bibfield  {title} {\bibinfo {title} {Information projections
  revisited},\ }\href {https://doi.org/10.1109/TIT.2003.810633} {\bibfield
  {journal} {\bibinfo  {journal} {IEEE Transactions on Information Theory}\
  }\textbf {\bibinfo {volume} {49}},\ \bibinfo {pages} {1474} (\bibinfo {year}
  {2003})}\BibitemShut {NoStop}%
\bibitem [{\citenamefont {Makhdoumi}\ \emph {et~al.}(2014)\citenamefont
  {Makhdoumi}, \citenamefont {Salamatian}, \citenamefont {Fawaz},\ and\
  \citenamefont {M{\'e}dard}}]{makhdoumi2014information}%
  \BibitemOpen
  \bibfield  {author} {\bibinfo {author} {\bibfnamefont {A.}~\bibnamefont
  {Makhdoumi}}, \bibinfo {author} {\bibfnamefont {S.}~\bibnamefont
  {Salamatian}}, \bibinfo {author} {\bibfnamefont {N.}~\bibnamefont {Fawaz}},\
  and\ \bibinfo {author} {\bibfnamefont {M.}~\bibnamefont {M{\'e}dard}},\
  }\bibfield  {title} {\bibinfo {title} {From the information bottleneck to the
  privacy funnel},\ }in\ \href@noop {} {\emph {\bibinfo {booktitle} {2014 IEEE
  Information Theory Workshop (ITW 2014)}}}\ (\bibinfo {organization} {IEEE},\
  \bibinfo {year} {2014})\ pp.\ \bibinfo {pages} {501--505}\BibitemShut
  {NoStop}%
\bibitem [{\citenamefont {Janzing}\ \emph {et~al.}(2013)\citenamefont
  {Janzing}, \citenamefont {Balduzzi}, \citenamefont {Grosse-Wentrup},\ and\
  \citenamefont {Sch{\"o}lkopf}}]{janzing2013quantifying}%
  \BibitemOpen
  \bibfield  {author} {\bibinfo {author} {\bibfnamefont {D.}~\bibnamefont
  {Janzing}}, \bibinfo {author} {\bibfnamefont {D.}~\bibnamefont {Balduzzi}},
  \bibinfo {author} {\bibfnamefont {M.}~\bibnamefont {Grosse-Wentrup}},\ and\
  \bibinfo {author} {\bibfnamefont {B.}~\bibnamefont {Sch{\"o}lkopf}},\
  }\bibfield  {title} {\bibinfo {title} {Quantifying causal influences},\
  }\href@noop {} {\bibfield  {journal} {\bibinfo  {journal} {The Annals of
  Statistics}\ }\textbf {\bibinfo {volume} {41}},\ \bibinfo {pages} {2324}
  (\bibinfo {year} {2013})}\BibitemShut {NoStop}%
\bibitem [{\citenamefont {Ay}(2021)}]{ay2021confounding}%
  \BibitemOpen
  \bibfield  {author} {\bibinfo {author} {\bibfnamefont {N.}~\bibnamefont
  {Ay}},\ }\bibfield  {title} {\bibinfo {title} {Confounding ghost channels and
  causality: a new approach to causal information flows},\ }\href@noop {}
  {\bibfield  {journal} {\bibinfo  {journal} {Vietnam journal of mathematics}\
  }\textbf {\bibinfo {volume} {49}},\ \bibinfo {pages} {547} (\bibinfo {year}
  {2021})}\BibitemShut {NoStop}%
\bibitem [{\citenamefont {Kolchinsky}\ and\ \citenamefont
  {Rocha}(2011)}]{kolchinsky2011prediction}%
  \BibitemOpen
  \bibfield  {author} {\bibinfo {author} {\bibfnamefont {A.}~\bibnamefont
  {Kolchinsky}}\ and\ \bibinfo {author} {\bibfnamefont {L.~M.}\ \bibnamefont
  {Rocha}},\ }\bibfield  {title} {\bibinfo {title} {Prediction and modularity
  in dynamical systems},\ }\href@noop {} {\bibfield  {journal} {\bibinfo
  {journal} {European Conference on Artificial Life (ECAL)}\ } (\bibinfo {year}
  {2011})}\BibitemShut {NoStop}%
\bibitem [{\citenamefont {Hidaka}\ and\ \citenamefont
  {Oizumi}(2018)}]{hidaka2018fast}%
  \BibitemOpen
  \bibfield  {author} {\bibinfo {author} {\bibfnamefont {S.}~\bibnamefont
  {Hidaka}}\ and\ \bibinfo {author} {\bibfnamefont {M.}~\bibnamefont
  {Oizumi}},\ }\bibfield  {title} {\bibinfo {title} {Fast and exact search for
  the partition with minimal information loss},\ }\href@noop {} {\bibfield
  {journal} {\bibinfo  {journal} {PLoS One}\ }\textbf {\bibinfo {volume}
  {13}},\ \bibinfo {pages} {e0201126} (\bibinfo {year} {2018})}\BibitemShut
  {NoStop}%
\bibitem [{\citenamefont {Rosas}\ \emph {et~al.}(2016)\citenamefont {Rosas},
  \citenamefont {Ntranos}, \citenamefont {Ellison}, \citenamefont {Pollin},\
  and\ \citenamefont {Verhelst}}]{rosas2016understanding}%
  \BibitemOpen
  \bibfield  {author} {\bibinfo {author} {\bibfnamefont {F.}~\bibnamefont
  {Rosas}}, \bibinfo {author} {\bibfnamefont {V.}~\bibnamefont {Ntranos}},
  \bibinfo {author} {\bibfnamefont {C.~J.}\ \bibnamefont {Ellison}}, \bibinfo
  {author} {\bibfnamefont {S.}~\bibnamefont {Pollin}},\ and\ \bibinfo {author}
  {\bibfnamefont {M.}~\bibnamefont {Verhelst}},\ }\bibfield  {title} {\bibinfo
  {title} {Understanding interdependency through complex information sharing},\
  }\href@noop {} {\bibfield  {journal} {\bibinfo  {journal} {Entropy}\ }\textbf
  {\bibinfo {volume} {18}},\ \bibinfo {pages} {38} (\bibinfo {year}
  {2016})}\BibitemShut {NoStop}%
\bibitem [{\citenamefont {Rosas}\ \emph {et~al.}(2019)\citenamefont {Rosas},
  \citenamefont {Mediano}, \citenamefont {Gastpar},\ and\ \citenamefont
  {Jensen}}]{rosas2019quantifying}%
  \BibitemOpen
  \bibfield  {author} {\bibinfo {author} {\bibfnamefont {F.~E.}\ \bibnamefont
  {Rosas}}, \bibinfo {author} {\bibfnamefont {P.~A.}\ \bibnamefont {Mediano}},
  \bibinfo {author} {\bibfnamefont {M.}~\bibnamefont {Gastpar}},\ and\ \bibinfo
  {author} {\bibfnamefont {H.~J.}\ \bibnamefont {Jensen}},\ }\bibfield  {title}
  {\bibinfo {title} {Quantifying high-order interdependencies via multivariate
  extensions of the mutual information},\ }\href@noop {} {\bibfield  {journal}
  {\bibinfo  {journal} {Physical Review E}\ }\textbf {\bibinfo {volume}
  {100}},\ \bibinfo {pages} {032305} (\bibinfo {year} {2019})}\BibitemShut
  {NoStop}%
\bibitem [{\citenamefont {Dubins}(1962)}]{dubins1962extreme}%
  \BibitemOpen
  \bibfield  {author} {\bibinfo {author} {\bibfnamefont {L.~E.}\ \bibnamefont
  {Dubins}},\ }\bibfield  {title} {\bibinfo {title} {On extreme points of
  convex sets},\ }\href@noop {} {\bibfield  {journal} {\bibinfo  {journal}
  {Journal of Mathematical Analysis and Applications}\ }\textbf {\bibinfo
  {volume} {5}},\ \bibinfo {pages} {237} (\bibinfo {year} {1962})}\BibitemShut
  {NoStop}%
\bibitem [{\citenamefont {Cover}\ and\ \citenamefont
  {Thomas}(2006)}]{cover_elements_2006}%
  \BibitemOpen
  \bibfield  {author} {\bibinfo {author} {\bibfnamefont {T.~M.}\ \bibnamefont
  {Cover}}\ and\ \bibinfo {author} {\bibfnamefont {J.~A.}\ \bibnamefont
  {Thomas}},\ }\href@noop {} {\emph {\bibinfo {title} {Elements of information
  theory}}}\ (\bibinfo  {publisher} {John Wiley \& Sons},\ \bibinfo {year}
  {2006})\BibitemShut {NoStop}%
\bibitem [{\citenamefont {Timo}\ \emph {et~al.}(2012)\citenamefont {Timo},
  \citenamefont {Grant},\ and\ \citenamefont {Kramer}}]{timo2012lossy}%
  \BibitemOpen
  \bibfield  {author} {\bibinfo {author} {\bibfnamefont {R.}~\bibnamefont
  {Timo}}, \bibinfo {author} {\bibfnamefont {A.}~\bibnamefont {Grant}},\ and\
  \bibinfo {author} {\bibfnamefont {G.}~\bibnamefont {Kramer}},\ }\bibfield
  {title} {\bibinfo {title} {Lossy broadcasting with complementary side
  information},\ }\href@noop {} {\bibfield  {journal} {\bibinfo  {journal}
  {IEEE Transactions on Information Theory}\ }\textbf {\bibinfo {volume}
  {59}},\ \bibinfo {pages} {104} (\bibinfo {year} {2012})}\BibitemShut
  {NoStop}%
\end{thebibliography}%

\end{document}